\documentclass[%
aps,
pra,%
amsmath,amssymb,
reprint,%
superscriptaddress,
floatfix
]{revtex4-1}

\usepackage{graphicx}
\usepackage{dcolumn}
\usepackage{bm}
\usepackage{
	times,mathptmx}
\usepackage[dvipsnames]{xcolor}
\usepackage{multirow}
\usepackage{enumitem}
\usepackage{chngpage}
\usepackage{float} 

\usepackage{amsmath}

\begin{document}
	
	\title{\bf Electrically empowered microcomb laser}
	
	\author{Jingwei Ling}
	\thanks{These authors contributed equally.}
	\affiliation{Department of Electrical and Computer Engineering, University of Rochester, Rochester, NY 14627, USA}
	\author{Zhengdong Gao}
	\thanks{These authors contributed equally.}
	\affiliation{Department of Electrical and Computer Engineering, University of Rochester, Rochester, NY 14627, USA}
	\author{Shixin Xue}
	\affiliation{Department of Electrical and Computer Engineering, University of Rochester, Rochester, NY 14627, USA}
	\author{Qili Hu}
	\affiliation{Institute of Optics, University of Rochester, Rochester, NY 14627, USA}
	\author{Mingxiao Li}
	\affiliation{Department of Electrical and Computer Engineering, University of Rochester, Rochester, NY 14627, USA}
	\author{Kaibo Zhang}
	\affiliation{Institute of Optics, University of Rochester, Rochester, NY 14627, USA}
	\author{Usman A. Javid}
	\affiliation{Institute of Optics, University of Rochester, Rochester, NY 14627, USA}
	\author{Raymond Lopez-Rios}
	\affiliation{Institute of Optics, University of Rochester, Rochester, NY 14627, USA}
	\author{Jeremy Staffa}
	\affiliation{Institute of Optics, University of Rochester, Rochester, NY 14627, USA}
	\author{Qiang Lin}
	\email{qiang.lin@rochester.edu}
	\affiliation{Department of Electrical and Computer Engineering, University of Rochester, Rochester, NY 14627, USA}
	\affiliation{Institute of Optics, University of Rochester, Rochester, NY 14627, USA}

	\begin{abstract}
			
			
			Optical frequency comb underpins a wide range of applications from communication, metrology, to sensing. Its development on a chip-scale platform -- so called soliton microcomb -- provides a promising path towards system miniaturization and functionality integration via photonic integrated circuit (PIC) technology. Although extensively explored in recent years, challenges remain in key aspects of microcomb such as complex soliton initialization, high threshold, low power efficiency, and limited comb reconfigurability. Here we present an on-chip laser that directly outputs microcomb and resolves all these challenges, with a distinctive mechanism created from synergetic interaction among resonant electro-optic effect, optical Kerr effect, and optical gain inside the laser cavity. Realized with integration between a III-V gain chip and a thin-film lithium niobate (TFLN) PIC, the laser is able to directly emit mode-locked microcomb on demand with robust turnkey operation inherently built in, with individual comb linewidth down to 600 Hz, whole-comb frequency tuning rate exceeding $\rm 2.4\times10^{17}$ Hz/s, and 100\% utilization of optical power fully contributing to comb generation. The demonstrated approach unifies architecture and operation simplicity, high-speed reconfigurability, and multifunctional capability enabled by TFLN PIC, opening up a great avenue towards on-demand generation of mode-locked microcomb that is expected to have profound impact on broad applications.\\

	\end{abstract}

	\maketitle
	\graphicspath{{Figures/}}

	Optical frequency comb is a coherent light source that consists of many highly coherent single-frequency laser lines equally spaced in the frequency domain. Its development has revolutionized many fields including metrology, spectroscopy, and clock \cite{diddams2020optical}. In recent years, significant interest has been attracted to the generation of phase-locked optical frequency comb in on-chip nonlinear microresonators \cite{kippenberg2018dissipative, gaeta2019photonic, chang2022integrated}. The superior coherence offered by these mode-locked microcombs has rendered a variety of important applications including data communication \cite{marin2017microresonator}, spectroscopic sensing \cite{yang2017microresonator}, optical computing \cite{xu202111, feldmann2021parallel}, range measurement \cite{suh2018soliton, trocha2018ultrafast, riemensberger2020massively, liu2020monolithic}, optical \cite{spencer2018optical} and microwave \cite{liang2015high} frequency synthesis, with many others expected in the years to come.  
	
	Despite these great progresses, challenges remain in the development and application of microcombs. The first is the difficulty in triggering comb mode-locking due to the intrinsic device nonlinearities. Recently, self-starting operation has been demonstrated to address this issue \cite{he2019self, shen2020integrated, bai2021brillouin, rowley2022self}. Their implementations, however, require sophisticated system pre-configuration and careful balance of specific nonlinear dynamics, which are difficult to apply in most practical devices. The second is the low power efficiency of soliton microcomb generation due to the pump-laser-cavity frequency detuning induced by soliton pulsing. Although pulse pumping \cite{obrzud2017temporal} or auxiliary-resonator enhancement \cite{xue2019super, bao2019laser} can improve the generation efficiency, they require delicate synchronization in time or resonance frequency and the difficulty of soliton initialization remains the same. The third is the limitation in the comb controllability due to the monolithic nature of the comb generator that is difficult to change after the device is fabricated. Piezoelectric effect could be used to deform the comb resonator \cite{liu2020monolithic}, which, however, exhibits limited tuning speed and efficiency due to its slow mechanical response. To date, the majority of comb generators still have to rely on external laser control to adjust the microcomb state. 
	
	Recently, there are significant advances in chip-scale integration of semiconductor lasers and nonlinear comb generators \cite{shen2020integrated, stern2018battery, raja2019electrically, xiang2021laser}, in which a diode laser produces single-frequency laser emission to pump a hybridly or heterogeneously integrated external nonlinear resonator to excite microcombs. Such a fully integrated system shows great promise in improving the size, weight, and power consumption. However, the nature of soliton comb generation remains essentially the same, with all the above challenges persistent. Up to now, the realization of an integrated comb source free from these challenges remains elusive. 
	
	\begin{figure*}
		\centering
		\includegraphics[width=\linewidth]{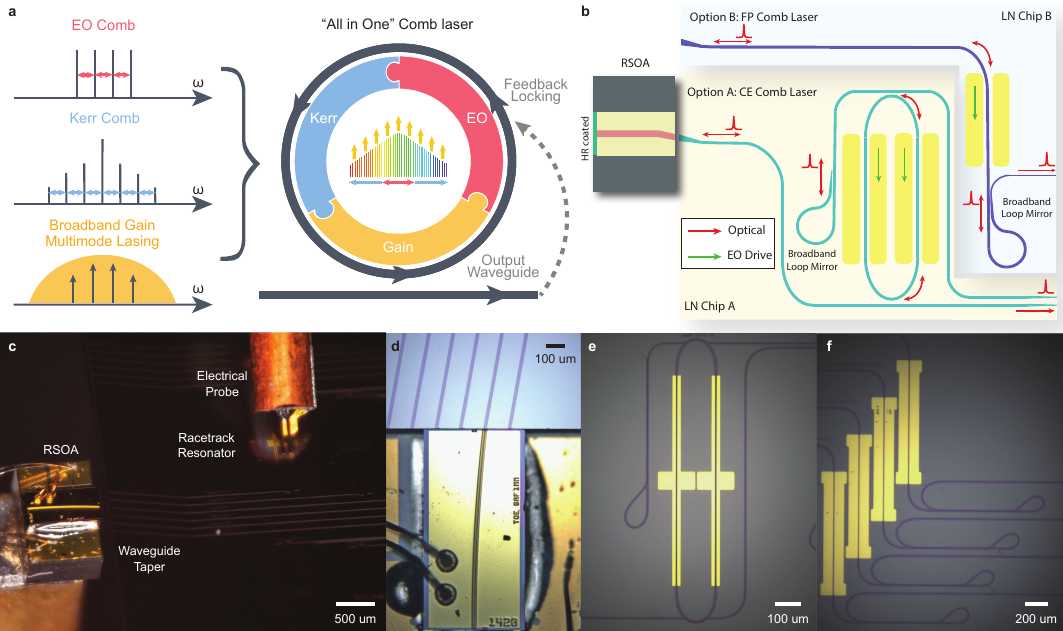}
		\caption{\label{Fig1} Device concept of the integrated comb laser. {\bf a.} Conceptual illustration of the comb generation and mode-locking principle, in which electro-optic (EO) comb generation, Kerr comb generation, and broadband optical gain all work synergistically together inside a single laser cavity for on-demand generation of mode-locked soliton comb. In addition, the laser comb output is detected and fed back to the laser cavity for resonant EO modulation to realize a self-sustained operation. {\bf b.} Schematic of comb laser cavity structure formed by hybrid integration between a RSOA chip and a LN external cavity chip. Two different configurations are employed: A, cavity-enhanced (CE) comb laser structure in which the LN external cavity is formed mainly by an embedded high-Q racetrack resonator together with a broadband Sagnacloop mirror; B, Fabry-Perot (FP) comb laser in which the LN external cavity is formed by an EO phase modulation section together with an a broadband Sagnac loop mirror. {\bf c.} Photo of a CE comb laser, showing that the RSOA is edge-coupled to the LN external cavity chip. {\bf d.} Zoom-in photo showing the edge-coupling region between the RSOA and the LN chip. {\bf e.}, Photo of the racetrack resonator and the loop mirror in a CE comb laser. {\bf f.} Photo of the EO phase modulator and the loop mirror in an FP comb laser.}
	\end{figure*}

	Here we present a fundamentally distinctive approach to resolve all these challenges in a single device. Figure \ref{Fig1}{\bf a} shows the device concept. In contrast to conventional approaches that rely solely on a single mechanism -- either optical Kerr or electro-optic effect -- for comb generation while with external pumping, we utilize the resonantly enhanced electro-optic (EO) modulation to initiate the comb generation, the resonantly enhanced optical Kerr effect to expand the comb bandwidth and phase-lock the comb lines, and the embedded III-V optical gain to sustain and stabilize the comb operation. Moreover, the resulting coherent microwave (via optical detection) is fed back to the EO comb to further enhance the mode-locking, leading to unique self-sustained comb operation. 

	We realize this approach by integrating a III-V gain element with a thin-film lithium-niobate (LN) photonic integrated circuit (PIC) to produce a III-V/LN comb laser (Fig.~\ref{Fig1}{\bf b}). LN PIC has attracted significant interest recently \cite{boes2018status, lin2020advances, zhu2021integrated, boes2023lithium} for a variety of applications including high-speed modulation \cite{wang2018integrated, he2019high}, frequency conversion \cite{wang2018ultrahigh, lu2020toward, mckenna2022ultra}, optical frequency comb \cite{he2019self, zhang2019broadband, gong2020near}, and single-frequency lasers \cite{de2021iii, han2021electrically, li2022integrated, ling2023self}. Here, we unite active EO modulation with passive four-wave mixing (FWM) in a dispersion-engineered high-Q laser cavity for the on-demand generation of mode-locked soliton microcomb, which naturally leads to self-starting full turnkey operation simply by turning on/off either the RF signal driving the comb resonator or the electric current driving the gain element. As the comb modes extract energy directly from material gain, 100\% of the optical power contributes directly to the comb generation. Moreover, the strong electro-optic effect of the LN cavity enables high tunability and reconfigurability of the produced microcomb. With this approach, we are able to produce broadband highly coherent microcombs, with individual comb linewidth down to 600 Hz, frequency tunability of over $\rm 2.4\times10^{17}$ Hz/s for the entire microcomb, microwave phase noise down to -115 dBc/Hz at 500 kHz frequency offset, and a wall-plug efficiency exceeding 5.6\%. The simplicity of the demonstrated approach opens up a new path for on-demand generation of mode-locked microcombs that is expected to have profound impact on the broad applications in high-precision metrology, telecommunications, remote sensing, clocking, computing, and beyond.

	\medskip
	\noindent{\sffamily\textbf{Comb laser structure design}}
	\medskip
	
	The III-V/LN microcomb laser is formed by integrating an InP reflective semiconductor optical amplifier (RSOA) with an LN external cavity chip via facet-to-facet coupling. We employ two types of laser cavity structures for the purpose, as shown in Fig.~\ref{Fig1}\textbf{b}. Chip A laser structure is embedded with a dispersion-engineered EO microresonator and Chip B laser structure consists of a simple EO phase modulation waveguide section. The advantage of Chip-A resonator-type structure is that the high-Q microresonator offers strong cavity enhancement for comb generation and mode-locking, which we term as the cavity-enhanced (CE) comb laser. The benefit of the Chip-B Fabry-Perot-type structure is that it offers flexibility of mode-locking operation, which we term as the Fabry-Perot (FP) comb laser. 
	
	\begin{figure*}[t]
		\centering
		\includegraphics[width=\linewidth]{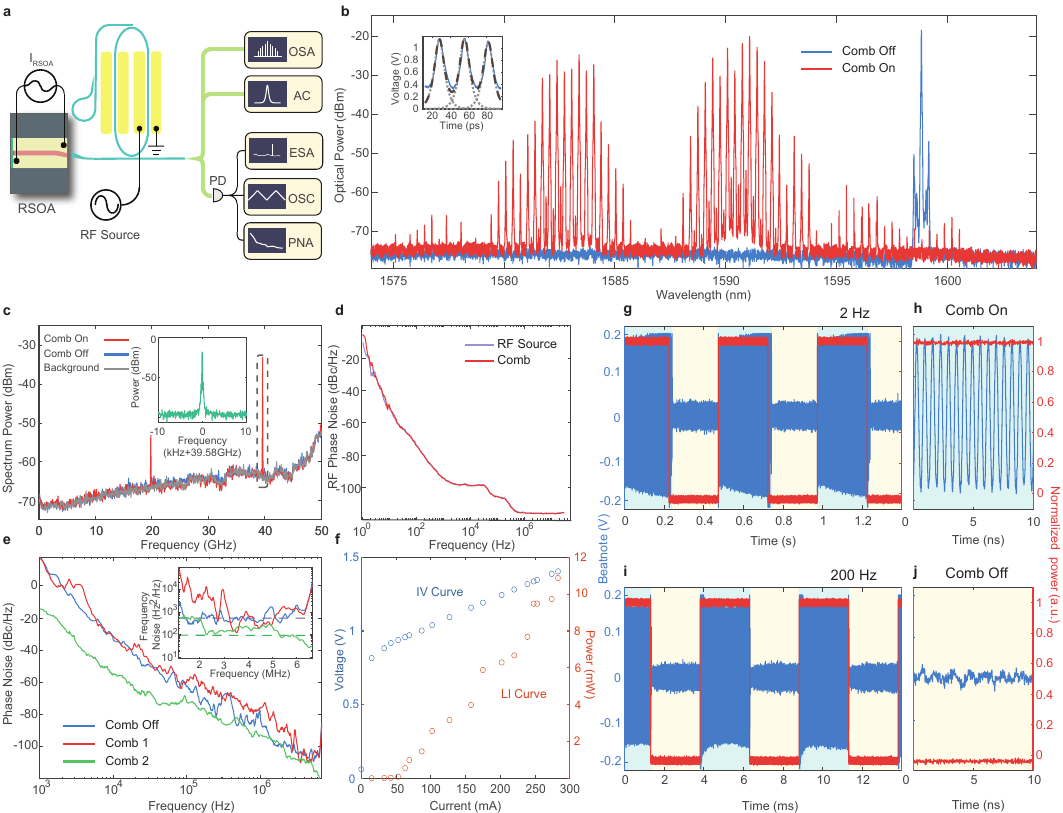}
		\caption{\label{Fig2} Lasing performance of a CE comb laser. {\bf a.} Schematic of the experimental setup for comb laser characterization. RSOA is powered by a stable current source and an RF signal generator is used to drive the racetrack resonator. OSA: optical spectrum analyzer; AC: autocorrelator; PD: photo-detector; ESA: electrical spectrum analyzer; OSC: real-time oscilloscope; PNA: phase noise analyzer. {\bf b.} Optical spectrum of the comb laser output. Off state: single mode lasing with the RF driving signal turned off. On state: comb lasing when the RF driving signal is turned on. Inset: Autocorrelation trace of the laser output pulses, in the "comb on" state, in which the blue curve shows the experimental data, dotted curves show the the fitted autocorrelation profiles from individual ${\rm sech^2}$ pulses, and the dashed curve show the overall fitted autocorrelation trace. The autocorrelation is recorded directly from the laser output pulses without dispersion compensation or pulse shaping. {\bf c.} Electrical spectrum of the beat note detected from the comb laser output. The red and blue curves show the comb-on and comb-off (single-mode lasing) states, respectively, corresponding to those in \textbf{b}. Gray curve shows the noise background of the optical detector, as a comparison. The inset shows the detailed spectrum of the RF beat note at 39.58~GHz. {\bf d.} Phase noise spectrum of the 39.58-GHz RF beat note (red) and the RF driving signal (blue). {\bf e.} Phase noise spectrum of the CE comb laser output measured with a self-heterodyne method. Red and green curves show for two different comb states, and blue curve shows for the comb-off (single-mode lasing) state. Inset shows the corresponding frequency noise spectrum of the three laser states. {\bf f.} P-I-V curve of the CE comb laser, in which the red and blue curves show the L-I and I-V curves, respectively. {\bf g - j.} Turnkey operation of the comb laser at two different speed of 2~Hz (\textbf{g, h}) and 200~Hz (\textbf{i, j}), respectively. Red curve shows the normalized driving RF power and blue curve shows the beating signal between the comb laser ouput and an external reference laser at 1582~nm. {\bf h} and {\bf j} show the zoomed-in signal for the on/off states, respectively. 
		}
	\end{figure*}

	In the CE comb laser, the group-velocity dispersion (GVD) of the racetrack microresonator (intrinsic optical Q $\sim 1.6\times 10^6$) is engineered to be small but slightly anomalous to support broadband comb generation. At the same time, the pulley coupling regions are specially designed for uniform close-to-critical coupling to the resonator over a broad telecom band. Such a design ensures high loaded optical Q ($\rm \sim 5 \times 10^5$) of the resonator uniformly across a wide spectral range which is crucial both for enhancing the comb generation and mode-locking and for efficient light coupling into/out of the microresonator. Moreover, we also engineer the GVD of straight waveguide sections outside the racetrack resonator to compensate for that of the RSOA section so as to minimize the overall laser cavity GVD. At the same time, the overall optical path length of the laser cavity is designed to be an integer multiple of the racetrack resonator's for matching their resonance mode frequencies and round-trip group delay. The GVD of the FP comb laser is engineered in a similar fashion. The free-spectral range (FSR) of the racetrack resonator is designed to be around 40~GHz to better accommodate bandwidths of RF filter and amplifier after detection of comb beating signal.\if{so that the beating between comb lines can be conveniently detected by an optical detector}\fi~In contrast, the FP comb laser is designed to have a small FSR of around 10~GHz for easy operation of harmonic mode locking.  
	
	\begin{figure*}
		\centering
		\includegraphics[width=\linewidth]{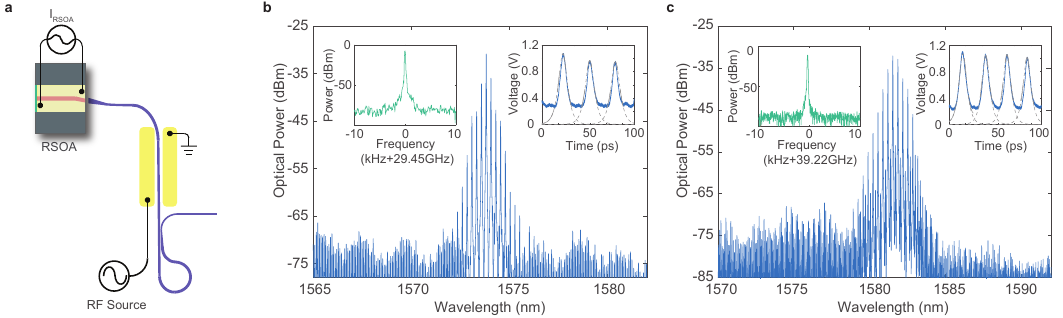}
		\caption{\label{Fig3} Harmonic mode-locking of a FP comb laser with an FSR of 9.817 GHz. Third harmonics (29.45 GHz) and fourth harmonics (39.27 GHz) are separately used as driving signals. \textbf{a.} Schematic of the experimental setup for harmonic mode-locking of the comb laser. \textbf{b.} Optical spectrum of the laser output with third-harmonic mode-locking, by driving the phase modulator with a RF signal at 29.45 GHz. \textbf{c.} Optical spectrum of the laser output with fourth-harmonic mode-locking, by driving the phase modulator with a RF signal at 39.27 GHz. In \textbf{b} and \textbf{c}, the left insets shows the electrical spectrum of the RF beat note detected from the output laser comb, and the right insets show the autocorrelation trace of the laser output pulses with dashed curve showing the fitted individual pulses. Same as Fig.~\ref{Fig2}\textbf{b}, autocorrelation is recorded directly from the laser output pulses without dispersion compensation or pulse shaping. 
		}
	\end{figure*}

	To support the broadband operation, the Sagnac loop mirror employs an adiabatic coupling design \cite{ramadan1998adiabatic} to achieve high reflection and feedback with a reflectivity of $>$95\% over a broad spectral band of 1500-1600 nm. On the other hand, a horn taper waveguide is designed on the LN chip to minimize the coupling loss between the LN chip and the RSOA gain chip. The RSOA exhibits a broadband gain in the telecom L-band, with a 3-dB bandwidth of $>$40~nm. Fig.~\ref{Fig1}\textbf{c}-\textbf{f} show the device structures. Details about the design parameters and the characterization of the laser structures are provided in the Supplementary Information (SI). 

	\medskip
	\noindent{\sffamily\textbf{Comb laser performance}}
	\medskip

	To excite the mode-locked microcomb, we first launch a single-frequency RF signal to drive the racetrack resonator of the CE comb laser (Fig.~\ref{Fig2}\textbf{a}), with a frequency of 39.58~GHz that matches its FSR. Before the RF signal is applied, the device exhibits single-mode or multi-mode lasing, with an example shown in the blue curve of Fig.~\ref{Fig2}\textbf{b}. However, A microcomb is readily produced as soon as the RF signal is applied, with an optical spectral bandwidth of about 20~nm (Fig.~\ref{Fig2}\textbf{b}, red curve). Mode-locking of the comb is verified by the clean RF tone at 39.58~GHz detected from the beating between comb lines (Fig.~\ref{Fig2}\textbf{c}), as well as the autocorrelation trace from the laser output pulses (Fig.~\ref{Fig2}\textbf{b}, inset). The 39.58-GHz RF tone exhibits a high signal-to-noise ratio of 79 dB  (Fig.~\ref{Fig2}\textbf{c}, inset), whose phase noise spectrum matches identically the driving RF source (Fig.~\ref{Fig2}\textbf{d}), showing the preservation of the relative phase coherence between comb lines via mode-locking. Mode-locking of the comb is also clearly evident by the clean noise floor around the DC region in the RF spectrum (Fig.~\ref{Fig2}\textbf{c}. See SI for details), where the zero extra noise from the mode-locked comb state infers that all the comb lines of the entire comb are phase-locked together. 
	
	The underlying mechanism responsible for mode-locking dominantly contributes from the combined resonant EO modulation and optical Kerr effect, in which the EO modulation produces EO sidebands to initiate the comb generation while the optical Kerr effect broadens the comb spectrum and phase-locks the comb lines (Fig.~\ref{Fig1}\textbf{a}). Indeed, the laser is able to produce mode-locked soliton pulses in the absence of EO modulation (while with a narrower spectrum), in which only the optical Kerr effect is responsible for mode-locking. The detailed theoretical modeling and testing results are provided in SI. In Fig.~\ref{Fig2}\textbf{c}, the small RF tone around the half-harmonic at 19.79 GHz indicates certain comb dynamics. It can be eliminated by reconfiguring the laser and one example is shown in SI which exhibits a clean single RF beating tone and a well-defined ${\rm sech^2}$-shaped soliton pulse spectrum. The two lasers mainly differ in their overall dispersion of the laser cavity, indicating that the device dispersion plays an important role on the comb spectrum. The two-sidelobe feature of the comb spectrum in Fig.~\ref{Fig2}\textbf{b} implies that the output pulses are likely to be mode-locked two-color pulses in which the two color pulses bounds with each other via certain interpulse interaction \cite{liao2020dual}. Its exact nature, however, will require further exploration. The comb spectrum can be reconfigured by changing the power or frequency of the RF driving signal, whose details are provided in the SI.

	\begin{figure*}[t]
		\centering
		\includegraphics[width=\linewidth]{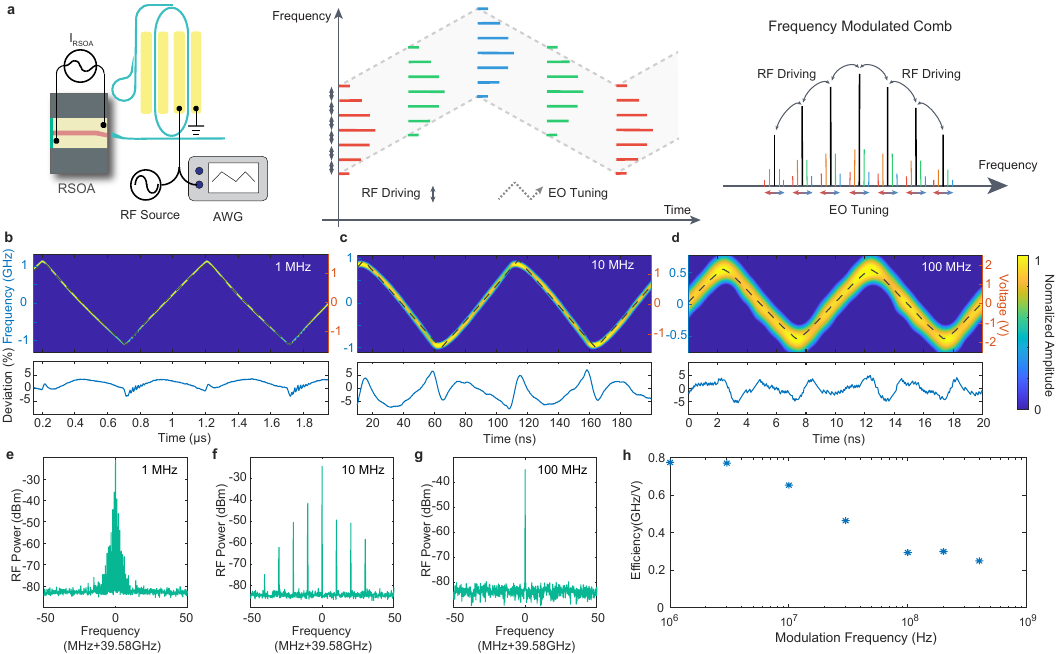}
		\caption{\label{Fig4} Fast frequency tuning of the whole lasing comb. \textbf{a.} Left panel: Schematic of the setup for comb frequency tuning, in which a triangular-waveform electrical signal produced by an arbitrary waveform generator (AWG) is used to drive the racetrack resonator of the CE comb laser together with the 39.58-GHz mode-locking RF signal. Middle panel: Conceptual illustration of the comb frequency tuning process, showing the laser frequencies of the comb are tuned together as a whole. Right panel: Schematic showing the corresponding sideband creation around the comb lines, introduced by triangular-waveform frequency modulation.  \textbf{b, c, d.} Time-frequency spectra of the beatnote between the comb laser output and a referenced laser operatoing at a fixed wavelength of 1582~nm, at the modulation speed of 1, 10, and 100~MHz, respectively. The dashed curves show the corresponding triangular-waveform EO tuning signal. Bottom panels: Corresponding relative frequency deviation at each modulation speed. \textbf{e, f, g.} Electrical spectrum of the 39.58-GHz beat note detected from the laser output comb, at modulation speed of 1 MHz (\textbf{e}), 10 MHz (\textbf{f}), and 100 MHz (\textbf{g}), respectively. \textbf{h.} Laser frequency tuning efficiency recorded at different modulation speeds. }
	\end{figure*}
	In addition to the high coherence between the comb lines, the comb laser also exhibits narrow linewidth on its individual comb lines. To show this feature, we employ the correlated self-heterodyne method \cite{camatel2008narrow, yuan2022correlated} to characterize the overall linewidth of the whole comb laser by launching the entire comb for linewidth measurement (rather than characterizing individual comb lines themselves) (See SI for details). The recorded phase noise spectrum is shown in Fig.~\ref{Fig2}\textbf{e}, which indicates a white frequency-noise floor of $\sim$350~$\rm Hz^2/Hz$ (Fig.~\ref{Fig2}\textbf{e}, inset) that corresponds to a laser linewidth of $\sim$2~kHz. The linewidth of the comb lines can be decreased further and an example is shown in Fig.~\ref{Fig2}\textbf{e} for a slightly different comb state produced from the same laser, which exhibits a white frequency-noise floor of $\sim$100~$\rm Hz^2/Hz$ (Fig.~\ref{Fig2}\textbf{e}, inset) that corresponds to a laser linewidth as low as $\sim$600~Hz. Note that these values represent the overall linewidth contributed from the entire comb, which indicates the upper limit of the intrinsic linewidth of individual comb lines. 
	
	Figure~\ref{Fig2}\textbf{f} shows the current-dependent characteristics of the comb laser, which exhibits a low threshold current of 50 mA, indicating the low overall loss of the integrated laser. The comb laser produces an optical output power of 11 mW at a pumping current of 275 mA and a pumping voltage of 1.4 V, which corresponds to a wall-plug efficiency of 2.8\%. As the laser has two output ports (Fig.~\ref{Fig1}\textbf{b}) that emit the same amount of optical power, the total wall-plug efficiency of the laser is thus 5.6\%. This level of wall-plug efficiency is on par with other integrated external-cavity semiconductor lasers recently developed \cite{xiang2021high, porter2023hybrid}. Intriguingly, the comb power increases with increased driving RF power, whose details are provided in the SI. Note that the total optical power contributes fully to the generated comb, in strong contrast to conventional Kerr solitons or EO combs in which the major optical power remains in the residual pump wave with low comb generation efficiency.  
	
	A distinctive feature of the comb laser is that the produced comb can be switched on/off on demand by simply switching on/off the driving RF signal. To show this feature, we beat the comb with a reference single-frequency laser operating at the wavelength of 1582~nm that is inside the comb spectrum, and monitor the beating signal with the RF driving signal being turned on/off. As shown in Fig.~\ref{Fig2}\textbf{g} and \textbf{i}, the beating signal follows faithfully the driving RF signal. The coherent beating signal shows up readily when the RF driving is on, indicating the generation of the mode-locked comb. The beating signal disappears right after the RF driving is off, indicating the shutoff of the comb state. Same phenomenon is observed when the reference laser is tuned to other wavelengths within the comb spectrum. 
		Similar phenomena are observed in the FP comb laser, while generally with smaller spectral extents due to the lack of cavity enhancement. The FP comb laser, however, exhibits a distinctive feature in that it can be flexibly mode-locked at higher harmonics of the laser cavity FSR. Fig.~\ref{Fig3} shows this feature. We are able to achieve third- and fourth-order harmonic mode-locking by applying an RF signal to the phase modulation section of the FP comb laser, with a frequency of 29.45 and 39.27~GHz, respectively, that are three and four times of the laser FSR (9.817~GHz). Again, mode locking of the combs is clearly verified by the detected RF beating signal from the combs with a SNR of 77 dB, as well as by the autocorrelation traces from the laser output pulses (Fig.~\ref{Fig3}\textbf{b},\textbf{c}, insets). 

		\begin{figure*}
			\centering
			\includegraphics[width=\linewidth]{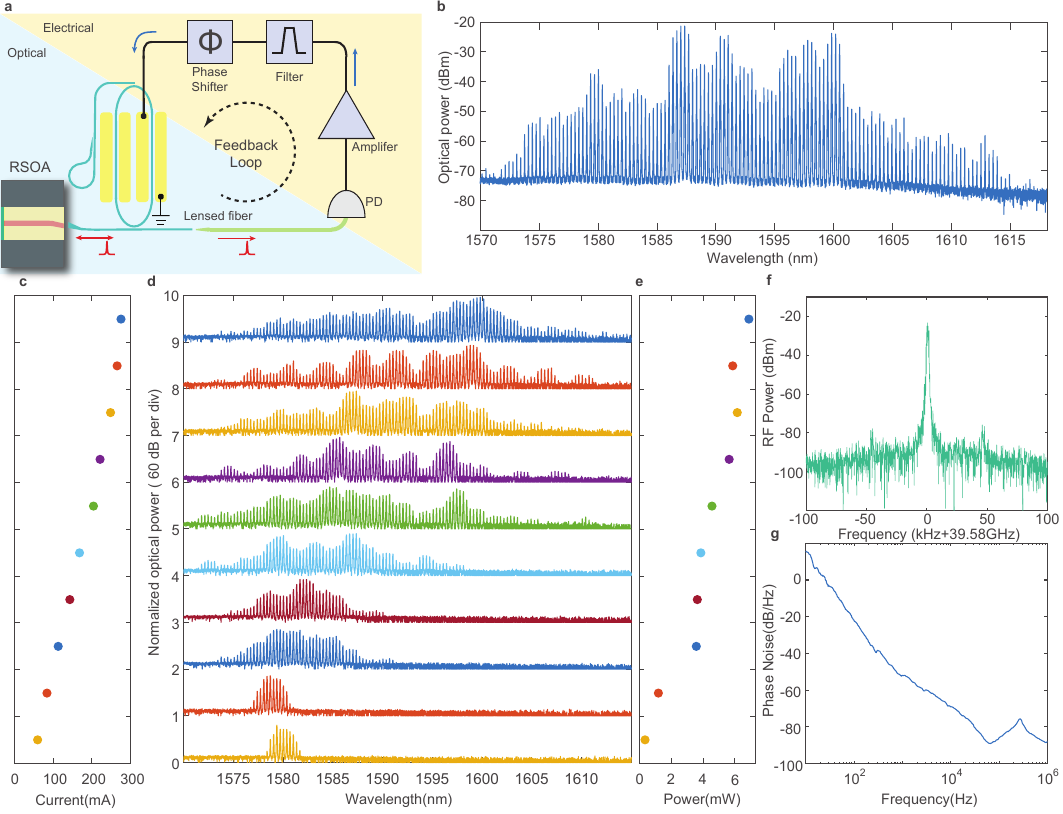}
			\caption{\label{Fig5} Self-sustained operation of the CE comb laser with feedback locking. \textbf{a.} Schematic of self-feedback locking of the CE comb laser.  \textbf{b.} Optical spectrum of the comb laser output at a driving current of 285 mA. \textbf{c - e.} Optical spectrum (\textbf{d}) and optical power (\textbf{e}) of the comb laser output as a function of the RSOA driving current (\textbf{c}). \textbf{f.} Electrical spectrum of the 39.58-GHz beat note detected from the laser output comb, with a driving current of 60 mA. \textbf{g.} Phase noise spectrum of the detected 39.58-GHz beat note. }
		\end{figure*}

		\medskip
		\noindent{\sffamily\textbf{High-speed frequency tuning of the comb laser}}
		\medskip
		
		
		Another distinctive characteristic of the comb laser is that the laser frequencies of the entire mode-locked comb can be tuned cohesively at an extremely high speed. To show this feature, we apply a triangular-waveform electric signal -- together with the 39.58-GHz RF driving signal -- to the racetrack resonator of the CE comb laser as shown in Fig.~\ref{Fig4}\textbf{a}. While the 39.58-GHz RF driving signal supports the mode-locking process, the triangular-waveform electric signal will adiabatically tune the resonance frequencies of the racetrack resonator, thus tuning the laser frequencies of the entire mode-locked comb together as a whole. 
		
		To show this feature, we beat the comb with a narrow-linewidth reference CW laser at 1582~nm that is about 15~GHz away from a comb line, and monitor the beating signal in real time. At the same time, we monitor the spectrum of the recorded 39.58-GHz RF tone from the beating between the comb lines (See SI for details of the setup). The frequency dynamics of the 15-GHz beating signal with the reference laser show the frequency tuning of the comb line nearby while the 39.58-GHz RF tone from the comb line beating indicates the quality of mode locking during the frequency tuning. Fig.~\ref{Fig4}\textbf{f}-\textbf{h} show the temporal variation of the 15-GHz beating signal at different modulation speeds of 1, 10, 100~MHz. They show clearly that the frequency tuning of the comb line follows faithfully the waveform of the driving triangular-waveform electric signal at all modulation speeds, with a deviation of no more than 5\%. In particular, the recorded 39.58-GHz RF tone from the comb line beating (Fig.~\ref{Fig4}\textbf{e}-\textbf{g}) remains unchanged during the frequency tuning, except with created modulation sidebands that simply results from the laser frequency modulation (see also Fig.~\ref{Fig4}\textbf{a}, right figure). This observation confirms that the phase-locking between the comb modes is fully preserved during the high-speed frequency tuning process, indicating that the entire mode-locked comb is tuned in its frequencies as a whole, without any perturbation to the comb mode spacing. This is in strong contrast to other comb modulation approaches \cite{liu2020monolithic, riemensberger2020massively, he2023high} where the comb mode spacing is seriously impacted by external modulation. The frequency tuning range of 1.2 GHz at the modulation speed of 100~MHz (Fig.~\ref{Fig4}\textbf{h}) corresponds to a frequency tuning rate as high as $\rm 2.4 \times 10^{17}  Hz/s$ for the comb. Both the frequency tuning rate and tuning speed are orders of magnitudes higher than the piezoelectric tuning and the external pump modulation approaches \cite{liu2020monolithic, riemensberger2020massively}, which are constrained only by the photon lifetime of the high-Q racetrack resonator. 
		As shown in Fig.~\ref{Fig4}\textbf{e}, the device exhibits a frequency tuning efficiency of about 0.2--0.8~GHz/V depending on the modulation speed, which is more than an order of magnitude higher than the piezoelectric approach \cite{liu2020monolithic}. The tuning efficiency can be further doubled by employing both sets of driving electrodes of the racetrack resonator (Fig.~\ref{Fig4}\textbf{a}).

		\medskip
		\noindent{\sffamily\textbf{Feedback mode-locking of the comb laser}}
		\medskip
		
		
		So far, the comb laser utilizes an external RF signal to support the mode-locking. This signal, however, can be removed by feeding the coherent 39.58-GHz RF tone detected from the comb mode beating directly back to the comb laser cavity to sustain the mode-locking, resulting in unique stand-alone self-sustained comb lasing operation. Figure~\ref{Fig5}\textbf{a} illustrates this approach. The comb laser output is detected by a high-speed optical detector whose output RF signal is amplified to an adequate amplitude, filtered to suppress excess low-frequency noises, adjusted with appropriate phase, and then fed back to drive the racetrack resonator of the CE comb laser. 
		
		As shown in Fig.~\ref{Fig5}\textbf{b}, a broadband microcomb with a spectrum covering about 50~nm and an optical power of 8.5~mW is produced on chip with a driving current of 285~mA. Indeed, the microcomb is readily produced on demand as soon as the driving current is turned on, with a driving current as low as 60~mA, as shown in Fig.~\ref{Fig5}\textbf{c-e}. Mode locking of the comb is clearly evident by the clean 39.58-GHz RF tone detected from the comb mode beating (Fig.~\ref{Fig5}\textbf{f}), which exhibits a SNR of 65 dB and a narrow 3-dB linewidth of 1.5 kHz. The phase noise of the RF beating signal reaches a level of -90 dBc/Hz at an offset frequency of 60 kHz, which is considerably lower than the laser heterodyne beating approach \cite{xiang20233d} and is comparable to that of free-running optical Kerr soliton microcombs \cite{liu2020photonic, wang2021towards, he2023high}. 
		The optical spectral bandwidth and the output power of the comb laser increase considerably with increased driving current (Fig.~\ref{Fig5}d). This is expected since the increased optical power of the mode-locked comb inside the high-Q racetrack resonator would significantly enhance the optical Kerr effect and the resulting four-wave-mixing process to broaden the comb spectrum. No saturation is observed on the comb spectral bandwidth as the current increases. 

		\medskip
		\noindent{\sffamily\textbf{Discussion}}\\
		\medskip
		
		The attainable extent of the microcomb spectrum or soliton pulse width in current devices is primarily limited by the available optical power inside the cavity and the group delay mismatch between the enhancing resonator and the main laser cavity. For the former, it can be improved by either reducing the loss (\emph{e.g.}, improving the RSOA-LN chip coupling efficiency) or increasing the optical gain (\emph{e.g.}, using a higher-power RSOA) inside the laser cavity. For the latter, our theoretical modeling (see SI) shows that the formation of ideal ultrashort soliton pulses would require that the roundtrip time of the main laser cavity be integer times that of the enhancing racetrack resonator. In current devices, however, there is a certain amount of mismatch which limits the comb spectrum and the coherence of the mode-beating RF tone. This problem can be resolved by further optimization of the roundtrip length of the main laser cavity and introducing tunability, after which we expect that ultra-broadband highly coherent soliton microcomb can be produced. 
		
		
		
		To conclude, we have introduced a new type of chip-scale microcomb laser that can be flexibly mode-locked with either active-driving or passive-feedback approaches and that can be tuned/reconfigured at an ultrafast speed, with robust turnkey operation inherently built in. 
		The demonstrated integrated comb laser exhibits outstanding reconfigurability and performance significantly beyond the reach of conventional on-chip mode-locked semiconductor lasers \cite{wang2017iii, davenport2018integrated, malinowski2019towards, hermans2022chip}. The demonstrated devices combine elegantly the simplicity of integrated laser structure, robustness of mode-locking operation, and electro-optically enhanced tunability and controllability, opening up a new avenue towards on-demand generation of soliton microcombs with high power efficiency that we envision to be of great promise for a wide range of applications including ranging, communication, optical and microwave synthesis, sensing, metrology, among many others. 

		\noindent\sffamily\textbf{Method}\\
		\noindent\rmfamily\textbf{Device fabrication}
		The device fabrication begins with a congruent x-cut thin film lithium-niobate-on-insulator (LNOI) wafer, with a 600 nm LN layer on a 4.7 $\rm \mu m$ silica-coated silicon substrate. E-beam lithography (EBL) and Ar-ion milling are used to etch the waveguide with ZEP-520A as mask. Etching thickness ranges from 300 nm (CE comb laser) to 350 nm (FP comb laser) for dispersion engineering. Second EBL is applied on PMMA for deposition of 400 nm gold-evaporated electrodes, which are placed 2.5 $\rm \mu m$ from the waveguide. The distance between the waveguide and electrode is chosen to balance the EO modulation frequency with loss from metal absorption. Dicing and polishing of LN chip are employed at last to acquire optimized fiber-to-chip and amplifier-to-chip coupling, with both coupling losses around 6 dB.\\

		\bibliography{all_reference.bib}

\providecommand{\noopsort}[1]{}\providecommand{\singleletter}[1]{#1}%
\begin{thebibliography}{10}
\expandafter\ifx\csname url\endcsname\relax
  \def\url#1{\texttt{#1}}\fi
\expandafter\ifx\csname urlprefix\endcsname\relax\def\urlprefix{URL }\fi
\providecommand{\bibinfo}[2]{#2}
\providecommand{\eprint}[2][]{\url{#2}}

\bibitem{diddams2020optical}
\bibinfo{author}{Diddams, S.~A.}, \bibinfo{author}{Vahala, K.} \&
  \bibinfo{author}{Udem, T.}
\newblock \bibinfo{title}{Optical frequency combs: Coherently uniting the
  electromagnetic spectrum}.
\newblock \emph{\bibinfo{journal}{Science}} \textbf{\bibinfo{volume}{369}},
  \bibinfo{pages}{eaay3676} (\bibinfo{year}{2020}).

\bibitem{kippenberg2018dissipative}
\bibinfo{author}{Kippenberg, T.~J.}, \bibinfo{author}{Gaeta, A.~L.},
  \bibinfo{author}{Lipson, M.} \& \bibinfo{author}{Gorodetsky, M.~L.}
\newblock \bibinfo{title}{Dissipative {Kerr} solitons in optical
  microresonators}.
\newblock \emph{\bibinfo{journal}{Science}} \textbf{\bibinfo{volume}{361}},
  \bibinfo{pages}{eaan8083} (\bibinfo{year}{2018}).

\bibitem{gaeta2019photonic}
\bibinfo{author}{Gaeta, A.~L.}, \bibinfo{author}{Lipson, M.} \&
  \bibinfo{author}{Kippenberg, T.~J.}
\newblock \bibinfo{title}{Photonic-chip-based frequency combs}.
\newblock \emph{\bibinfo{journal}{Nature Photonics}}
  \textbf{\bibinfo{volume}{13}}, \bibinfo{pages}{158--169}
  (\bibinfo{year}{2019}).

\bibitem{chang2022integrated}
\bibinfo{author}{Chang, L.}, \bibinfo{author}{Liu, S.} \&
  \bibinfo{author}{Bowers, J.~E.}
\newblock \bibinfo{title}{Integrated optical frequency comb technologies}.
\newblock \emph{\bibinfo{journal}{Nature Photonics}}
  \textbf{\bibinfo{volume}{16}}, \bibinfo{pages}{95--108}
  (\bibinfo{year}{2022}).

\bibitem{marin2017microresonator}
\bibinfo{author}{Marin-Palomo, P.} \emph{et~al.}
\newblock \bibinfo{title}{Microresonator-based solitons for massively parallel
  coherent optical communications}.
\newblock \emph{\bibinfo{journal}{Nature}} \textbf{\bibinfo{volume}{546}},
  \bibinfo{pages}{274--279} (\bibinfo{year}{2017}).

\bibitem{yang2017microresonator}
\bibinfo{author}{Suh, M.-G.}, \bibinfo{author}{Yang, Q.-F.},
  \bibinfo{author}{Yang, K.~Y.}, \bibinfo{author}{Yi, X.} \&
  \bibinfo{author}{Vahala, K.~J.}
\newblock \bibinfo{title}{Microresonator soliton dual-comb spectroscopy}.
\newblock \emph{\bibinfo{journal}{Science}} \textbf{\bibinfo{volume}{354}},
  \bibinfo{pages}{600--603} (\bibinfo{year}{2016}).

\bibitem{xu202111}
\bibinfo{author}{Xu, X.} \emph{et~al.}
\newblock \bibinfo{title}{{11 TOPS photonic convolutional accelerator for
  optical neural networks}}.
\newblock \emph{\bibinfo{journal}{Nature}} \textbf{\bibinfo{volume}{589}},
  \bibinfo{pages}{44--51} (\bibinfo{year}{2021}).

\bibitem{feldmann2021parallel}
\bibinfo{author}{Feldmann, J.} \emph{et~al.}
\newblock \bibinfo{title}{Parallel convolutional processing using an integrated
  photonic tensor core}.
\newblock \emph{\bibinfo{journal}{Nature}} \textbf{\bibinfo{volume}{589}},
  \bibinfo{pages}{52--58} (\bibinfo{year}{2021}).

\bibitem{suh2018soliton}
\bibinfo{author}{Suh, M.-G.} \& \bibinfo{author}{Vahala, K.~J.}
\newblock \bibinfo{title}{Soliton microcomb range measurement}.
\newblock \emph{\bibinfo{journal}{Science}} \textbf{\bibinfo{volume}{359}},
  \bibinfo{pages}{884--887} (\bibinfo{year}{2018}).

\bibitem{trocha2018ultrafast}
\bibinfo{author}{Trocha, P.} \emph{et~al.}
\newblock \bibinfo{title}{Ultrafast optical ranging using microresonator
  soliton frequency combs}.
\newblock \emph{\bibinfo{journal}{Science}} \textbf{\bibinfo{volume}{359}},
  \bibinfo{pages}{887--891} (\bibinfo{year}{2018}).

\bibitem{riemensberger2020massively}
\bibinfo{author}{Riemensberger, J.} \emph{et~al.}
\newblock \bibinfo{title}{Massively parallel coherent laser ranging using a
  soliton microcomb}.
\newblock \emph{\bibinfo{journal}{Nature}} \textbf{\bibinfo{volume}{581}},
  \bibinfo{pages}{164--170} (\bibinfo{year}{2020}).

\bibitem{liu2020monolithic}
\bibinfo{author}{Liu, J.} \emph{et~al.}
\newblock \bibinfo{title}{Monolithic piezoelectric control of soliton
  microcombs}.
\newblock \emph{\bibinfo{journal}{Nature}} \textbf{\bibinfo{volume}{583}},
  \bibinfo{pages}{385--390} (\bibinfo{year}{2020}).

\bibitem{spencer2018optical}
\bibinfo{author}{Spencer, D.~T.} \emph{et~al.}
\newblock \bibinfo{title}{An optical-frequency synthesizer using integrated
  photonics}.
\newblock \emph{\bibinfo{journal}{Nature}} \textbf{\bibinfo{volume}{557}},
  \bibinfo{pages}{81--85} (\bibinfo{year}{2018}).

\bibitem{liang2015high}
\bibinfo{author}{Liang, W.} \emph{et~al.}
\newblock \bibinfo{title}{{High spectral purity Kerr frequency comb radio
  frequency photonic oscillator}}.
\newblock \emph{\bibinfo{journal}{Nature Communications}}
  \textbf{\bibinfo{volume}{6}}, \bibinfo{pages}{7957} (\bibinfo{year}{2015}).

\bibitem{he2019self}
\bibinfo{author}{He, Y.} \emph{et~al.}
\newblock \bibinfo{title}{{Self-starting bi-chromatic {LiNbO$_3$} soliton
  microcomb}}.
\newblock \emph{\bibinfo{journal}{Optica}} \textbf{\bibinfo{volume}{6}},
  \bibinfo{pages}{1138--1144} (\bibinfo{year}{2019}).

\bibitem{shen2020integrated}
\bibinfo{author}{Shen, B.} \emph{et~al.}
\newblock \bibinfo{title}{Integrated turnkey soliton microcombs}.
\newblock \emph{\bibinfo{journal}{Nature}} \textbf{\bibinfo{volume}{582}},
  \bibinfo{pages}{365--369} (\bibinfo{year}{2020}).

\bibitem{bai2021brillouin}
\bibinfo{author}{Bai, Y.} \emph{et~al.}
\newblock \bibinfo{title}{{Brillouin-Kerr soliton frequency combs in an optical
  microresonator}}.
\newblock \emph{\bibinfo{journal}{Physical Review Letters}}
  \textbf{\bibinfo{volume}{126}}, \bibinfo{pages}{063901}
  (\bibinfo{year}{2021}).

\bibitem{rowley2022self}
\bibinfo{author}{Rowley, M.} \emph{et~al.}
\newblock \bibinfo{title}{Self-emergence of robust solitons in a microcavity}.
\newblock \emph{\bibinfo{journal}{Nature}} \textbf{\bibinfo{volume}{608}},
  \bibinfo{pages}{303--309} (\bibinfo{year}{2022}).

\bibitem{obrzud2017temporal}
\bibinfo{author}{Obrzud, E.}, \bibinfo{author}{Lecomte, S.} \&
  \bibinfo{author}{Herr, T.}
\newblock \bibinfo{title}{Temporal solitons in microresonators driven by
  optical pulses}.
\newblock \emph{\bibinfo{journal}{Nature Photonics}}
  \textbf{\bibinfo{volume}{11}}, \bibinfo{pages}{600--607}
  (\bibinfo{year}{2017}).

\bibitem{xue2019super}
\bibinfo{author}{Xue, X.}, \bibinfo{author}{Zheng, X.} \&
  \bibinfo{author}{Zhou, B.}
\newblock \bibinfo{title}{Super-efficient temporal solitons in mutually coupled
  optical cavities}.
\newblock \emph{\bibinfo{journal}{Nature Photonics}}
  \textbf{\bibinfo{volume}{13}}, \bibinfo{pages}{616--622}
  (\bibinfo{year}{2019}).

\bibitem{bao2019laser}
\bibinfo{author}{Bao, H.} \emph{et~al.}
\newblock \bibinfo{title}{Laser cavity-soliton microcombs}.
\newblock \emph{\bibinfo{journal}{Nature Photonics}}
  \textbf{\bibinfo{volume}{13}}, \bibinfo{pages}{384--389}
  (\bibinfo{year}{2019}).

\bibitem{stern2018battery}
\bibinfo{author}{Stern, B.}, \bibinfo{author}{Ji, X.},
  \bibinfo{author}{Okawachi, Y.}, \bibinfo{author}{Gaeta, A.~L.} \&
  \bibinfo{author}{Lipson, M.}
\newblock \bibinfo{title}{Battery-operated integrated frequency comb
  generator}.
\newblock \emph{\bibinfo{journal}{Nature}} \textbf{\bibinfo{volume}{562}},
  \bibinfo{pages}{401--405} (\bibinfo{year}{2018}).

\bibitem{raja2019electrically}
\bibinfo{author}{Raja, A.~S.} \emph{et~al.}
\newblock \bibinfo{title}{Electrically pumped photonic integrated soliton
  microcomb}.
\newblock \emph{\bibinfo{journal}{Nature Communications}}
  \textbf{\bibinfo{volume}{10}}, \bibinfo{pages}{680} (\bibinfo{year}{2019}).

\bibitem{xiang2021laser}
\bibinfo{author}{Xiang, C.} \emph{et~al.}
\newblock \bibinfo{title}{Laser soliton microcombs heterogeneously integrated
  on silicon}.
\newblock \emph{\bibinfo{journal}{Science}} \textbf{\bibinfo{volume}{373}},
  \bibinfo{pages}{99--103} (\bibinfo{year}{2021}).

\bibitem{boes2018status}
\bibinfo{author}{Boes, A.}, \bibinfo{author}{Corcoran, B.},
  \bibinfo{author}{Chang, L.}, \bibinfo{author}{Bowers, J.} \&
  \bibinfo{author}{Mitchell, A.}
\newblock \bibinfo{title}{{Status and potential of lithium niobate on insulator
  (LNOI) for photonic integrated circuits}}.
\newblock \emph{\bibinfo{journal}{Laser \& Photonics Reviews}}
  \textbf{\bibinfo{volume}{12}}, \bibinfo{pages}{1700256}
  (\bibinfo{year}{2018}).

\bibitem{lin2020advances}
\bibinfo{author}{Lin, J.}, \bibinfo{author}{Bo, F.}, \bibinfo{author}{Cheng,
  Y.} \& \bibinfo{author}{Xu, J.}
\newblock \bibinfo{title}{Advances in on-chip photonic devices based on lithium
  niobate on insulator}.
\newblock \emph{\bibinfo{journal}{Photonics Research}}
  \textbf{\bibinfo{volume}{8}}, \bibinfo{pages}{1910--1936}
  (\bibinfo{year}{2020}).

\bibitem{zhu2021integrated}
\bibinfo{author}{Zhu, D.} \emph{et~al.}
\newblock \bibinfo{title}{Integrated photonics on thin-film lithium niobate}.
\newblock \emph{\bibinfo{journal}{Advances in Optics and Photonics}}
  \textbf{\bibinfo{volume}{13}}, \bibinfo{pages}{242--352}
  (\bibinfo{year}{2021}).

\bibitem{boes2023lithium}
\bibinfo{author}{Boes, A.} \emph{et~al.}
\newblock \bibinfo{title}{Lithium niobate photonics: Unlocking the
  electromagnetic spectrum}.
\newblock \emph{\bibinfo{journal}{Science}} \textbf{\bibinfo{volume}{379}},
  \bibinfo{pages}{eabj4396} (\bibinfo{year}{2023}).

\bibitem{wang2018integrated}
\bibinfo{author}{Wang, C.} \emph{et~al.}
\newblock \bibinfo{title}{{Integrated lithium niobate electro-optic modulators
  operating at CMOS-compatible voltages}}.
\newblock \emph{\bibinfo{journal}{Nature}} \textbf{\bibinfo{volume}{562}},
  \bibinfo{pages}{101--104} (\bibinfo{year}{2018}).

\bibitem{he2019high}
\bibinfo{author}{He, M.} \emph{et~al.}
\newblock \bibinfo{title}{{High-performance hybrid silicon and lithium niobate
  Mach--Zehnder modulators for 100 Gbit s${^{-1}}$ and beyond}}.
\newblock \emph{\bibinfo{journal}{Nature Photonics}}
  \textbf{\bibinfo{volume}{13}}, \bibinfo{pages}{359--364}
  (\bibinfo{year}{2019}).

\bibitem{wang2018ultrahigh}
\bibinfo{author}{Wang, C.} \emph{et~al.}
\newblock \bibinfo{title}{Ultrahigh-efficiency wavelength conversion in
  nanophotonic periodically poled lithium niobate waveguides}.
\newblock \emph{\bibinfo{journal}{Optica}} \textbf{\bibinfo{volume}{5}},
  \bibinfo{pages}{1438--1441} (\bibinfo{year}{2018}).

\bibitem{lu2020toward}
\bibinfo{author}{Lu, J.}, \bibinfo{author}{Li, M.}, \bibinfo{author}{Zou,
  C.-L.}, \bibinfo{author}{Al~Sayem, A.} \& \bibinfo{author}{Tang, H.~X.}
\newblock \bibinfo{title}{Toward 1\% single-photon anharmonicity with
  periodically poled lithium niobate microring resonators}.
\newblock \emph{\bibinfo{journal}{Optica}} \textbf{\bibinfo{volume}{7}},
  \bibinfo{pages}{1654--1659} (\bibinfo{year}{2020}).

\bibitem{mckenna2022ultra}
\bibinfo{author}{McKenna, T.~P.} \emph{et~al.}
\newblock \bibinfo{title}{Ultra-low-power second-order nonlinear optics on a
  chip}.
\newblock \emph{\bibinfo{journal}{Nature Communications}}
  \textbf{\bibinfo{volume}{13}}, \bibinfo{pages}{4532} (\bibinfo{year}{2022}).

\bibitem{zhang2019broadband}
\bibinfo{author}{Zhang, M.} \emph{et~al.}
\newblock \bibinfo{title}{Broadband electro-optic frequency comb generation in
  a lithium niobate microring resonator}.
\newblock \emph{\bibinfo{journal}{Nature}} \textbf{\bibinfo{volume}{568}},
  \bibinfo{pages}{373--377} (\bibinfo{year}{2019}).

\bibitem{gong2020near}
\bibinfo{author}{Gong, Z.}, \bibinfo{author}{Liu, X.}, \bibinfo{author}{Xu, Y.}
  \& \bibinfo{author}{Tang, H.~X.}
\newblock \bibinfo{title}{Near-octave lithium niobate soliton microcomb}.
\newblock \emph{\bibinfo{journal}{Optica}} \textbf{\bibinfo{volume}{7}},
  \bibinfo{pages}{1275--1278} (\bibinfo{year}{2020}).

\bibitem{de2021iii}
\bibinfo{author}{de~Beeck, C.~O.} \emph{et~al.}
\newblock \bibinfo{title}{{III/V-on-lithium niobate amplifiers and lasers}}.
\newblock \emph{\bibinfo{journal}{Optica}} \textbf{\bibinfo{volume}{8}},
  \bibinfo{pages}{1288--1289} (\bibinfo{year}{2021}).

\bibitem{han2021electrically}
\bibinfo{author}{Han, Y.} \emph{et~al.}
\newblock \bibinfo{title}{{Electrically pumped widely tunable O-band hybrid
  lithium niobite/III-V laser}}.
\newblock \emph{\bibinfo{journal}{Optics Letters}}
  \textbf{\bibinfo{volume}{46}}, \bibinfo{pages}{5413--5416}
  (\bibinfo{year}{2021}).

\bibitem{li2022integrated}
\bibinfo{author}{Li, M.} \emph{et~al.}
\newblock \bibinfo{title}{{Integrated Pockels laser}}.
\newblock \emph{\bibinfo{journal}{Nature Communications}}
  \textbf{\bibinfo{volume}{13}}, \bibinfo{pages}{5344} (\bibinfo{year}{2022}).

\bibitem{ling2023self}
\bibinfo{author}{Ling, J.} \emph{et~al.}
\newblock \bibinfo{title}{Self-injection locked frequency conversion laser}.
\newblock \emph{\bibinfo{journal}{Laser \& Photonics Reviews}}
  \bibinfo{pages}{2200663} (\bibinfo{year}{2023}).

\bibitem{ramadan1998adiabatic}
\bibinfo{author}{Ramadan, T.~A.} \& \bibinfo{author}{Osgood, R.~M.}
\newblock \bibinfo{title}{Adiabatic couplers: design rules and optimization}.
\newblock \emph{\bibinfo{journal}{Journal of lightwave technology}}
  \textbf{\bibinfo{volume}{16}}, \bibinfo{pages}{277} (\bibinfo{year}{1998}).

\bibitem{liao2020dual}
\bibinfo{author}{Liao, R.} \emph{et~al.}
\newblock \bibinfo{title}{{Dual-comb generation from a single laser source:
  principles and spectroscopic applications towards mid-IR—A review}}.
\newblock \emph{\bibinfo{journal}{Journal of Physics: Photonics}}
  \textbf{\bibinfo{volume}{2}}, \bibinfo{pages}{042006} (\bibinfo{year}{2020}).

\bibitem{camatel2008narrow}
\bibinfo{author}{Camatel, S.} \& \bibinfo{author}{Ferrero, V.}
\newblock \bibinfo{title}{{Narrow linewidth CW laser phase noise
  characterization methods for coherent transmission system applications}}.
\newblock \emph{\bibinfo{journal}{Journal of Lightwave Technology}}
  \textbf{\bibinfo{volume}{26}}, \bibinfo{pages}{3048--3055}
  (\bibinfo{year}{2008}).

\bibitem{yuan2022correlated}
\bibinfo{author}{Yuan, Z.} \emph{et~al.}
\newblock \bibinfo{title}{Correlated self-heterodyne method for ultra-low-noise
  laser linewidth measurements}.
\newblock \emph{\bibinfo{journal}{Optics Express}}
  \textbf{\bibinfo{volume}{30}}, \bibinfo{pages}{25147--25161}
  (\bibinfo{year}{2022}).

\bibitem{xiang2021high}
\bibinfo{author}{Xiang, C.} \emph{et~al.}
\newblock \bibinfo{title}{High-performance silicon photonics using
  heterogeneous integration}.
\newblock \emph{\bibinfo{journal}{IEEE Journal of Selected Topics in Quantum
  Electronics}} \textbf{\bibinfo{volume}{28}}, \bibinfo{pages}{1--15}
  (\bibinfo{year}{2021}).

\bibitem{porter2023hybrid}
\bibinfo{author}{Porter, C.}, \bibinfo{author}{Zeng, S.},
  \bibinfo{author}{Zhao, X.} \& \bibinfo{author}{Zhu, L.}
\newblock \bibinfo{title}{Hybrid integrated chip-scale laser systems}.
\newblock \emph{\bibinfo{journal}{APL Photonics}} \textbf{\bibinfo{volume}{8}}
  (\bibinfo{year}{2023}).

\bibitem{he2023high}
\bibinfo{author}{He, Y.} \emph{et~al.}
\newblock \bibinfo{title}{High-speed tunable microwave-rate soliton microcomb}.
\newblock \emph{\bibinfo{journal}{Nature Communications}}
  \textbf{\bibinfo{volume}{14}}, \bibinfo{pages}{3467} (\bibinfo{year}{2023}).

\bibitem{xiang20233d}
\bibinfo{author}{Xiang, C.} \emph{et~al.}
\newblock \bibinfo{title}{3d integration enables ultralow-noise isolator-free
  lasers in silicon photonics}.
\newblock \emph{\bibinfo{journal}{Nature}} \textbf{\bibinfo{volume}{620}},
  \bibinfo{pages}{78--85} (\bibinfo{year}{2023}).

\bibitem{liu2020photonic}
\bibinfo{author}{Liu, J.} \emph{et~al.}
\newblock \bibinfo{title}{Photonic microwave generation in the {X}-and {K}-band
  using integrated soliton microcombs}.
\newblock \emph{\bibinfo{journal}{Nature Photonics}}
  \textbf{\bibinfo{volume}{14}}, \bibinfo{pages}{486--491}
  (\bibinfo{year}{2020}).

\bibitem{wang2021towards}
\bibinfo{author}{Wang, B.} \emph{et~al.}
\newblock \bibinfo{title}{Towards high-power, high-coherence, integrated
  photonic mmwave platform with microcavity solitons}.
\newblock \emph{\bibinfo{journal}{Light: Science \& Applications}}
  \textbf{\bibinfo{volume}{10}}, \bibinfo{pages}{4} (\bibinfo{year}{2021}).

\bibitem{wang2017iii}
\bibinfo{author}{Wang, Z.} \emph{et~al.}
\newblock \bibinfo{title}{{A III-V-on-Si ultra-dense comb laser}}.
\newblock \emph{\bibinfo{journal}{Light: Science \& Applications}}
  \textbf{\bibinfo{volume}{6}}, \bibinfo{pages}{e16260--e16260}
  (\bibinfo{year}{2017}).

\bibitem{davenport2018integrated}
\bibinfo{author}{Davenport, M.~L.}, \bibinfo{author}{Liu, S.} \&
  \bibinfo{author}{Bowers, J.~E.}
\newblock \bibinfo{title}{{Integrated heterogeneous silicon/III--V mode-locked
  lasers}}.
\newblock \emph{\bibinfo{journal}{Photonics Research}}
  \textbf{\bibinfo{volume}{6}}, \bibinfo{pages}{468--478}
  (\bibinfo{year}{2018}).

\bibitem{malinowski2019towards}
\bibinfo{author}{Malinowski, M.} \emph{et~al.}
\newblock \bibinfo{title}{Towards on-chip self-referenced frequency-comb
  sources based on semiconductor mode-locked lasers}.
\newblock \emph{\bibinfo{journal}{Micromachines}}
  \textbf{\bibinfo{volume}{10}}, \bibinfo{pages}{391} (\bibinfo{year}{2019}).

\bibitem{hermans2022chip}
\bibinfo{author}{Hermans, A.}, \bibinfo{author}{Van~Gasse, K.} \&
  \bibinfo{author}{Kuyken, B.}
\newblock \bibinfo{title}{On-chip optical comb sources}.
\newblock \emph{\bibinfo{journal}{APL Photonics}} \textbf{\bibinfo{volume}{7}}
  (\bibinfo{year}{2022}).

\end{thebibliography}


\providecommand{\noopsort}[1]{}\providecommand{\singleletter}[1]{#1}%
\begin{thebibliography}{1}
\expandafter\ifx\csname url\endcsname\relax
  \def\url#1{\texttt{#1}}\fi
\expandafter\ifx\csname urlprefix\endcsname\relax\def\urlprefix{URL }\fi
\providecommand{\bibinfo}[2]{#2}
\providecommand{\eprint}[2][]{\url{#2}}

\bibitem{herr2014mode}
\bibinfo{author}{Herr, T.} \emph{et~al.}
\newblock \bibinfo{title}{Mode spectrum and temporal soliton formation in
  optical microresonators}.
\newblock \emph{\bibinfo{journal}{Physical review letters}}
  \textbf{\bibinfo{volume}{113}}, \bibinfo{pages}{123901}
  (\bibinfo{year}{2014}).

\bibitem{yi2015soliton}
\bibinfo{author}{Yi, X.}, \bibinfo{author}{Yang, Q.-F.}, \bibinfo{author}{Yang,
  K.~Y.}, \bibinfo{author}{Suh, M.-G.} \& \bibinfo{author}{Vahala, K.}
\newblock \bibinfo{title}{Soliton frequency comb at microwave rates in a
  high-{Q} silica microresonator}.
\newblock \emph{\bibinfo{journal}{Optica}} \textbf{\bibinfo{volume}{2}},
  \bibinfo{pages}{1078--1085} (\bibinfo{year}{2015}).

\bibitem{ramadan1998adiabatic}
\bibinfo{author}{Ramadan, T.~A.} \& \bibinfo{author}{Osgood, R.~M.}
\newblock \bibinfo{title}{Adiabatic couplers: design rules and optimization}.
\newblock \emph{\bibinfo{journal}{Journal of lightwave technology}}
  \textbf{\bibinfo{volume}{16}}, \bibinfo{pages}{277} (\bibinfo{year}{1998}).

\bibitem{camatel2008narrow}
\bibinfo{author}{Camatel, S.} \& \bibinfo{author}{Ferrero, V.}
\newblock \bibinfo{title}{Narrow linewidth cw laser phase noise
  characterization methods for coherent transmission system applications}.
\newblock \emph{\bibinfo{journal}{Journal of Lightwave Technology}}
  \textbf{\bibinfo{volume}{26}}, \bibinfo{pages}{3048--3055}
  (\bibinfo{year}{2008}).

\bibitem{yuan2022correlated}
\bibinfo{author}{Yuan, Z.} \emph{et~al.}
\newblock \bibinfo{title}{Correlated self-heterodyne method for ultra-low-noise
  laser linewidth measurements}.
\newblock \emph{\bibinfo{journal}{Optics Express}}
  \textbf{\bibinfo{volume}{30}}, \bibinfo{pages}{25147--25161}
  (\bibinfo{year}{2022}).

\bibitem{tran2019tutorial}
\bibinfo{author}{Tran, M.~A.}, \bibinfo{author}{Huang, D.} \&
  \bibinfo{author}{Bowers, J.~E.}
\newblock \bibinfo{title}{Tutorial on narrow linewidth tunable semiconductor
  lasers using si/iii-v heterogeneous integration}.
\newblock \emph{\bibinfo{journal}{APL photonics}} \textbf{\bibinfo{volume}{4}}
  (\bibinfo{year}{2019}).

\bibitem{coen2013universal}
\bibinfo{author}{Coen, S.} \& \bibinfo{author}{Erkintalo, M.}
\newblock \bibinfo{title}{Universal scaling laws of kerr frequency combs}.
\newblock \emph{\bibinfo{journal}{Optics letters}}
  \textbf{\bibinfo{volume}{38}}, \bibinfo{pages}{1790--1792}
  (\bibinfo{year}{2013}).

\bibitem{zhang2019broadband}
\bibinfo{author}{Zhang, M.} \emph{et~al.}
\newblock \bibinfo{title}{Broadband electro-optic frequency comb generation in
  a lithium niobate microring resonator}.
\newblock \emph{\bibinfo{journal}{Nature}} \textbf{\bibinfo{volume}{568}},
  \bibinfo{pages}{373--377} (\bibinfo{year}{2019}).

\end{thebibliography}
		\bibliographystyle{naturemag} %
		\noindent\sffamily\textbf{Acknowledgements}
		
		\noindent\rmfamily 	The authors would like to thank Lin Chang and William Renninger for helpful discussions. This work is supported in part by the Defense Advanced Research Projects Agency (DARPA) QuICC program under Agreement No.~FA8650-23-C-7312 and LUMOS program under Agreement No.~HR001-20-2-0044, and the National Science Foundation (NSF) under Grant No.~OMA-2138174 and ECCS-2231036. This work was performed in part at the Cornell NanoScale Facility, a member of the National Nanotechnology Coordinated Infrastructure (National Science Foundation, ECCS-1542081); and at the Cornell Center for Materials Research (National Science Foundation, Grant No.~DMR-1719875). \newline

		\noindent\sffamily\textbf{Author Contributions}\nolinebreak
		
		\noindent\rmfamily J.L. and Z.G. designed and fabricated the devices. J.L. and Z.G. performed the device characterization. S.X and Q.H. assisted in the device fabrication. S.X.,M.L., K.Z., U.J., R.L. and J.S. assisted in experiments. J.L., Z.G., and Q.L. wrote the manuscript with contributions from all authors. Q.L. supervised the project. Q.L. conceived the concept. \newline
		
		\noindent\sffamily\textbf{Additional Informations}\\
		\noindent\rmfamily\textbf{Supplementary information} Supplementary Information is available for this paper.\\
		\noindent\textbf{Competing interests.} The authors declare no competing interests.  Correspondence and requests for materials should be sent to  Q.L. (qiang.lin@rochester.edu).\\
		\noindent\rmfamily\textbf{Reprints and permission.} Reprints and permissions are available at www.nature.com/reprints.\\
	\end{document}


\title{\bf Supplementary Information for\\{"Electrically enpowered micro-comb laser"}}
\author{Jingwei Ling}
\thanks{These authors contributed equally.}
\affiliation{Department of Electrical and Computer Engineering, University of Rochester, Rochester, NY 14627, USA}
\author{Zhengdong Gao}
\thanks{These authors contributed equally.}
\affiliation{Department of Electrical and Computer Engineering, University of Rochester, Rochester, NY 14627, USA}
\author{Shixin Xue}
\affiliation{Department of Electrical and Computer Engineering, University of Rochester, Rochester, NY 14627, USA}
\author{Qili Hu}
\affiliation{Institute of Optics, University of Rochester, Rochester, NY 14627, USA}
\author{Mingxiao Li}
\affiliation{Department of Electrical and Computer Engineering, University of Rochester, Rochester, NY 14627, USA}
\author{Kaibo Zhang}
\affiliation{Institute of Optics, University of Rochester, Rochester, NY 14627, USA}
\author{Usman A. Javid}
\affiliation{Institute of Optics, University of Rochester, Rochester, NY 14627, USA}
\author{Raymond Lopez-Rios}
\affiliation{Institute of Optics, University of Rochester, Rochester, NY 14627, USA}
\author{Jeremy Staffa}
\affiliation{Institute of Optics, University of Rochester, Rochester, NY 14627, USA}
\author{Qiang Lin}
\email{qiang.lin@rochester.edu}
\affiliation{Department of Electrical and Computer Engineering, University of Rochester, Rochester, NY 14627, USA}
\affiliation{Institute of Optics, University of Rochester, Rochester, NY 14627, USA}

\begin{abstract}
In this supplementary material, we provide detailed information on the comb laser design, characterization, data processing, and theoretical modeling. 

\end{abstract}
\maketitle
\graphicspath{{Figures/}}

\section{LN Device Characterization}
\begin{figure*}
	\centering
	\includegraphics[width=\linewidth]{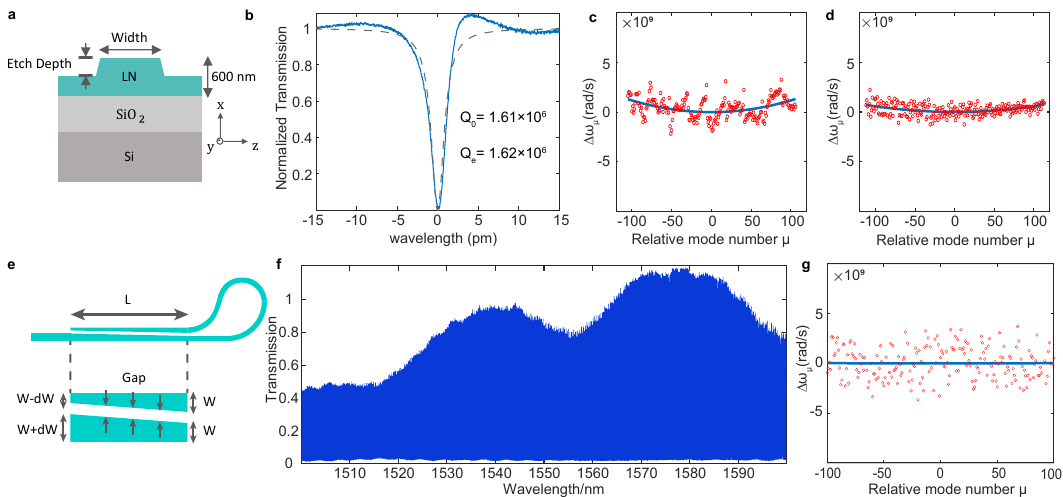}
	\caption{\label{FigS4} Individual components in the comb lasers and their passive performance. \textbf{a}, Schematic for the cross-section of the LN waveguide. \textbf{b}, Transmission curve of racetrack resonator with a width of 1.6 $\mu m$, critically coupled with a single bus waveguide. The dashed grey fitting curve corresponds to an intrinsic Q of 1.6 million. \textbf{c} and \textbf{d}, Mode dispersion curve versus the relative mode number, $\mu$, for $\rm TE_{00}$ mode of resonator $\alpha$ and $\beta$. The red circles show the $\rm \Delta\omega_{\mu}$ of each mode and the blue curves show the parabolic fittings. The GVD $\beta_2$ are fitted to be $-0.02~{\rm ps^2/m}$ for 1.6-$\mu m$-wide racetrack $\alpha$ in \textbf{c}, and $-0.01~{\rm ps^2/m}$ for 1.5-$\mu m$-wide racetrack $\beta$ in \textbf{d}. \textbf{e}, Schematic of the adiabatic Sagnac loop mirror. \textbf{f}, Reflection spectrum from loop mirror measurement. \if{FP resonance is formed between the Sagnac mirror and the polished waveguide facet.}\fi \textbf{g}, Mode dispersion curve for the $\rm TE_{00}$ mode of total FP laser cavity, showing a $\beta_2$ of $-0.003~{\rm ps^2/m}$.}
\end{figure*}

Our LN PIC is based on a directly etched waveguide structure depicted in Fig.~\ref{FigS4}\textbf{a}, with LN layer thickness of 600~nm. The CE comb laser and the FP comb laser have slightly different etching depths for dispersion management (300~nm for the CE comb laser, and 350~nm for the FP comb laser). Due to the difficulty of measuring the separate properties of each component within the CE comb laser cavity, we fabricated and tested the individual components such as racetrack resonators and Sagnac mirrors on the same wafer using identical fabrication parameters.
\par In this section, we will introduce the detailed device geometry and characterize the passive properties of the individual components, which include the optical Q of racetrack resonators, the reflection spectrum of Sagnac mirror, the dispersion of racetrack resonator and that of the main laser cavities. All the measurements are done in the telecom band. 
\subsection{Racetrack resonator}
Here, we show the measurement results of the racetrack resonator for two different CE comb laser devices $\alpha$ and $\beta$. The device $\alpha$ with a racetrack waveguide wide of 1.6 ${\rm \mu m}$ is used for the results shown in Fig.~2, Fig.~4, and Fig.~5 in the main text. Another device $\beta$ with a waveguide wide of 1.5 ${\rm \mu m}$ 
is shown as a comparison which will be further investigated in the following section.

\par We individually tested the transmission of the racetrack resonators with an external telecom tunable laser. The first parameter to characterize is the optical Q of the racetracks. Measuring the transmission trace of a critical coupled racetrack resonator with a width of 1.6 $\rm \mu m$, we obtained an intrinsic Q of 1.6 million as shown in Fig.~\ref{FigS4}\textbf{b}. Optical Q at a similar level is obtained for the 1.5-$\rm \mu m$ racetrack $\beta$. We then characterized the dispersion of the racetrack resonators by measuring the cavity resonance frequency $\omega_\mu$ within a wavelength range from 1500 nm to 1600 nm, around the reference frequency $\omega_0$ that corresponds to the wavelength at 1550 nm \cite{herr2014mode,yi2015soliton}. The cavity resonance frequency follows the relationship, $\omega_\mu = \omega_0 + \mu D_1 + \frac{1}{2} \mu^2 D_2 + \frac{1}{6} \mu^3 D_3 + \cdots$, where $D_1/(2\pi)$ is the FSR around $\omega_0$ and $D_2$ is related to the group velocity dispersion (GVD) parameter ${\beta_2}$ as $D_2 = -\frac{c}{n} D_1^2 \beta_2$. 
Figure~\ref{FigS4}\textbf{c} and \textbf{d} shows the experimentally recorded $\Delta\omega_\mu = \omega_\mu - \omega_0$ for Device $\alpha$ and $\beta$ respectively, from which we obtained ${\beta_2} = -0.02~{\rm ps^2/m}$ for the 1.6-${\rm \mu m}$ racetrack $\alpha$ and ${\beta_2} = -0.01~{\rm ps^2/m}$ for the 1.5-$\mu m$ racetrack $\beta$. 

\subsection{Sagnac mirror}
The schematic of the adiabatically coupled Sagnac mirror is depicted in Fig.~\ref{FigS4}\textbf{e}. We adopt different Sagnac mirror designs for different comb laser devices based on the dispersion and reflectance requirements. A typical Sagnac mirror used for the CE comb laser in the main text has a long adiabatic coupling section\cite{ramadan1998adiabatic} (L = 600 $\mu m$), with 300-nm etching depth, 0.8-$\rm \mu m$ width W and 0.3-$\rm \mu m$ gap between the waveguide. An adiabatic smooth transition to the waveguide of $\rm W \pm dW = 0.9/0.7 \mu m$ is designed to be the coupling region. This adiabatic transition at the coupling region enables a broadband 50\%:50\% power splitting for the coupler and near 100\% broadband reflection for the Sagnac loop mirror. To test the device, we butt-coupled light into a stand-alone Sagnac mirror and measured its reflection spectrum which is shown in Fig.~\ref{FigS4}\textbf{f}. The recorded reflection spectrum shows strong oscillation which simply results from the Fabry-Perot-type resonance formed between the waveguide input facet and the Sagnac mirror. The consistently high and uniform extinction ratio serves as a compelling evidence of the Sagnac mirror's broadband high reflection. The gradual wavelength dependence of the spectrum amplitude is due to factors related to testing, such as wavelength-dependence of the input optical power, slightly dispersive reflection, and the limited bandwidth of optical circulator. 

\subsection{Laser cavity}
Dispersion of the comb laser cavity matters for comb generation, especially for the FP comb laser. The average dispersion of Sagnac mirror together with bus waveguides and RSOA is designed to be near zero. The near-zero dispersion facilitates comb generation through both EO modulation and FWM. This can also be characterized by measuring the FSR of resonances from the reflected oscillating spectrum. The fitting curve of $\Delta\omega_\mu$ is shown in Fig.~\ref{FigS4}\textbf{g}, for the FP comb laser device used in the main text. Due to the larger linewidth of the FP cavity resonance, slightly larger fluctuation is observed on the spectrum. Nevertheless, the averaged dispersion parameter $\beta_2$ is characterized to be $-0.003~{\rm ps^{2}/m}$, a value that is close to zero.\\
\indent On the other hand, CE comb lasers contain two cavities, the racetrack resonator, and the main laser cavity, the former of which dominates the comb generation process, with dispersion characterized in the previous section. We also minimize the average dispersion of the main laser cavity by designing the Sagnac mirror, and bus waveguides. 

\section{Laser performance analysis}
In this section, we provide additional information regarding the performance of the comb lasers. We will show first alternate comb states, followed by two different CE comb lasers' behavior under varying RF driving signals. Then we will show the comb laser device working at a passive mode-locking state, with no EO driving. Furthermore, this section will incorporate details not addressed in the main text, including the linewidth measurements of comb and the approach adopted for comb tuning and data processing.
\subsection{Performance of different comb lasers}
The typical comb lasing states of device $\alpha$ are demonstrated in Fig.2 of the main text (marked as $\alpha$). We also designed another comb laser (marked as $\beta$) with different dispersion and tested its comb lasing performance. The design differences between these two devices have been elaborated in the previous section.\\
\indent We show here the performance of the device $\beta$ with an FSR of 43.90 GHz. The comb outputs from the two different devices differ in the spectrum: Laser $\alpha$ features a comb output with two sidelobes; while Laser $\beta$ has a single envelope with a $sech^2$ shape. 
\begin{figure}
	\centering
	\includegraphics[width=\linewidth]{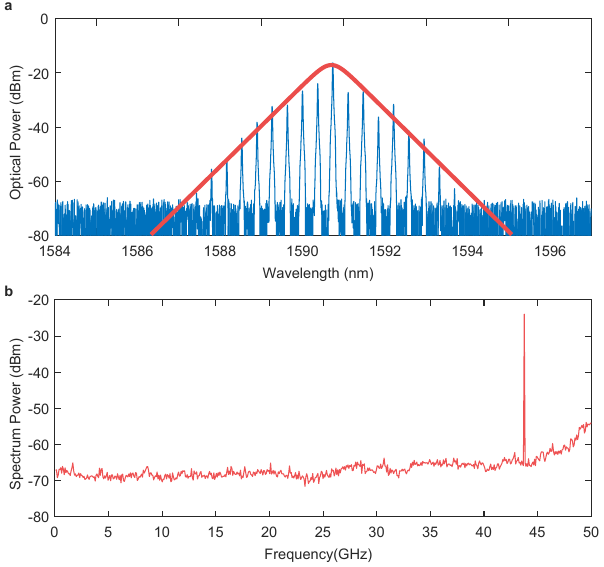}
	\caption{\label{single_envelope} Optical comb generated from Laser $\beta$. \textbf{a}, Optical spectrum of the generated comb,  The blue comb spectrum is fitted with red $sech^2$ curve. \textbf{b}, Electrical spectrum of the beat note detected from the comb laser output, with single tone at one FSR. DC background or half-harmonic tone is not observed in the spectrum.}
\end{figure}
As can be seen from Fig.~\ref{single_envelope}\textbf{a}, the $sech^2$ comb has a bandwidth of 7 nm. The spectrum of the comb-line beating note given in Fig.~\ref{single_envelope}\textbf{b} shows a clear single beating tone at 43.9 GHz with no extra tone or DC noise, showcasing the clear phase locking. 
These findings align well with the theoretical prediction, which will be discussed in Fig.~\ref{FigS2_1} of the following section. Noticeably, half-harmonic RF line doesn't appear in this case which indicates that it may relate to the unknown dynamics in Laser $\alpha$.
\subsection{Reconfiguration of comb lasers}
\begin{figure*}
\centering
\includegraphics[width=\linewidth]{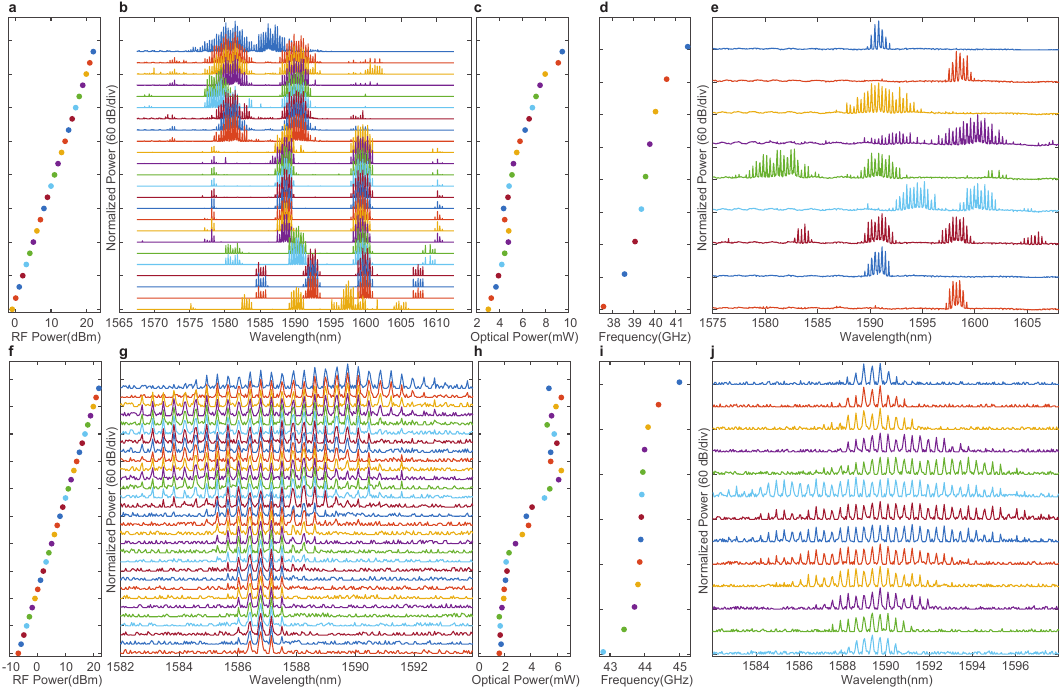}
\caption{\label{Fig3} Reconfiguration of CE comb lasers. Comb spectra at different RF powers and frequencies are demonstrated, for device $\alpha$ (FSR=39.58 GHz) in \textbf{a}-\textbf{e} and for device $\beta$ (FSR= 43.90 GHz) in \textbf{f}-\textbf{j}. \textbf{a}-\textbf{c}, Performances of Laser $\alpha$ with 39.58 GHz RF driving but different RF powers: \textbf{a}, Input RF driving power; \textbf{b}, Normalized comb output spectra; \textbf{c}, Output optical power, growing with increased RF input. \textbf{d}-\textbf{e}, Performances of Laser $\alpha$ with 20 dBm RF driving at different frequencies: \textbf{d}, Input RF driving frequency; \textbf{e}, Normalized comb output spectra at different driving frequencies, bandwidth maximizes at one FSR. \textbf{f}-\textbf{h}, Performances of Laser $\beta$ with 43.90 GHz RF driving but different RF powers: \textbf{f}, Input RF driving power; \textbf{g}, Normalized comb output spectra; \textbf{h}, Output optical power with increased RF input, growing with increased RF driving and saturating at 6mW. \textbf{i}-\textbf{j}, Performances of Laser $\beta$ with 20 dBm RF driving at different frequencies: \textbf{i}, Input RF driving frequency; \textbf{j}, Normalized comb output spectra at different driving frequencies, bandwidth maximizes at one FSR.}
\end{figure*}

To further investigate the dynamics and performance, and to show the reconfigurability of the comb laser, we tested the performance of CE comb laser $\alpha$ and $\beta$ under different RF driving signals. The optical bandwidths and mode-lock states can be adjusted with different electrical inputs. With the CE comb laser device $\alpha$ shown in the main text, we fixed the current at 270 mA and measured the optical spectra of the comb at different driving RF powers and RF frequencies. As evident from Fig.~\ref{Fig3}\textbf{b}-\textbf{c}, increasing the power of the RF driving signal leads to an expansion of the comb spectrum and a corresponding rise in output power. Of particular interest is that when the driving current is held constant, the RF driving and comb bandwidth influence not only the comb power of the laser but also the output of RSOA itself. Such an intriguing enhancement effect accounts partially for the high wall-plug efficiency we mentioned in the main text.\\
\indent The repetition rate and comb FSR can be slightly tuned by varying the driving frequency. To quantitatively characterize this effect, we examined Laser $\alpha$ output spectrum versus the driving frequency deviation from the racetrack resonator's FSR. An increase in the detuning of the driving RF frequency from the resonator FSR leads to a reduction in output comb bandwidth, as evident in Fig.~\ref{Fig3}\textbf{e}. Specifically for Laser $\alpha$, greatly increased detuning alters the comb shape from two sidelobes to a single lobe. \\
\indent The reconfigurability of the comb laser is further verified with the other CE comb laser $\beta$. Although the two devices differ in design, and thus different dispersion and comb shapes, they exhibit consistent trends in response to the changes in the RF driving power and frequency. The consistencies are proved by Fig.~\ref{Fig3}\textbf{f}-\textbf{h} and Fig.~\ref{Fig3}\textbf{i}-\textbf{j}, for RF power and frequency, respectively.\\ 
\indent Sharing the similar comb evolution behavior with driving RF signals change, the two laser devices are still drastically different in the comb spectra. The varied dispersion and cavity roundtrips may be the primary contributors to the different comb generation in Fig.~\ref{Fig3}. In Laser $\beta$, we notice the comb output spectra are more localized with narrower spectra bandwidths, while the comb spectral envelopes of Laser $\alpha$ tend to shift around with different RF drivings. One possible explanation is that better group delay matching is achieved between the racetrack resonator and main laser cavity for device $\alpha$, which allows for the formation of a broader range of comb lines at different wavelengths. However, the details of the mechanism behind require further study.
\subsection{Passive mode locking}
\begin{figure}
	\centering
	\includegraphics[width=0.8\linewidth]{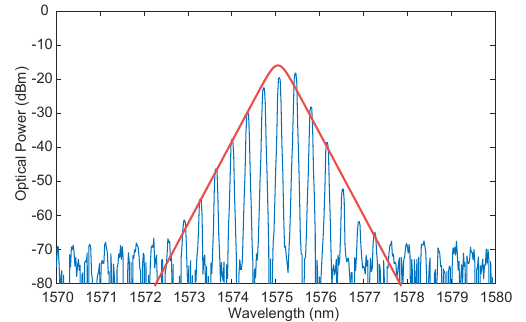}
	\caption{\label{passive} Optical comb generated from passive mode locking in CE comb laser, with no EO modulation applied. The blue comb spectrum is fitted with red $sech^2$ curve.}
\end{figure}

The comb laser is also able to operate without the EO driving or initiation 
in which the Kerr-based FWM serves as the mode-locking mechanism. 
With another CE comb laser (marked as $\gamma$) which has a racetrack resonator with lower coupling loss and higher loaded Q, we observed such passive mode locking as shown in Fig.~\ref{passive}. Despite a narrower bandwidth than that shown in Fig.~\ref{single_envelope}, the comb spectrum also features a $sech^2$ spectral shape as shown by the fitted red curve. The comb state also exhibits a clean electrical comb-line beating tone and time-domain pulses measured by autocorrelation (not shown), as an indication of phase-locking. The spectral bandwidth of the passive mode-locked comb is mainly limited by the RSOA output power and the roundtrip mismatch issue we described in the discussion section of the main text.\\
\subsection{Spectrum of the comb-line beat note around the DC region}
\begin{figure}
	\centering
	\includegraphics[width=0.8\linewidth]{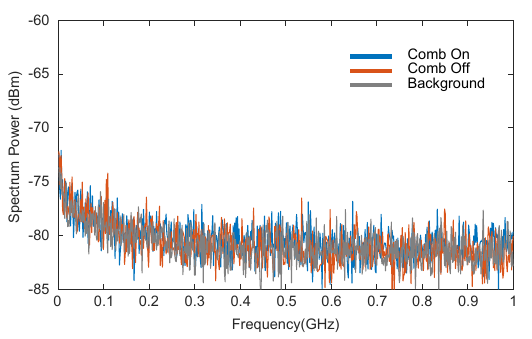}
	\caption{\label{DC} Zoom-in Electrical spectrum of the CE comb laser output beating signal, ranging from DC to 1 GHz. }
\end{figure}

In the main text, we have shown the comprehensive spectrum of the comb-line beating node over a broad bandwidth for the comb produced by the CE comb laser $\alpha$. Here, we present a detailed view of this spectrum, for the frequency region from DC to 1 GHz. For comparison, we also include in Fig.~\ref{DC} both spectra of the background noise originating from the photodetector and of the single-mode lasing state. Remarkably, these three curves exhibit a complete overlap. The low noise around the DC region serves as a compelling evidence of effective phase locking that exists among all the comb lines.

\subsection{Comb linewidth measurement}
\begin{figure}
	\centering
	\includegraphics[width=\linewidth]{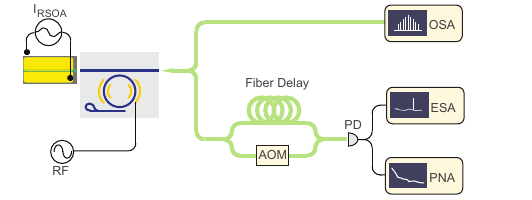}
	\caption{\label{Fig_linewidth} Experimental setup for measuring the linewidth of the comb. Sub-coherent length self-heterodyne method is used. The frequency is shifted by 55 MHz with an AOM , and single-mode fiber delays with 30 meter are used for heterodyne. AOM: acoustic optical modulator; OSA: optical spectrum analyzer; PD: photo-detector; ESA: electrical spectrum analyzer; PNA: phase noise analyzer.}
\end{figure}
The linewidth measurement is performed using the sub-coherent length self-heterodyne method \cite{camatel2008narrow,yuan2022correlated,tran2019tutorial}. The setup is shown in Fig.~\ref{Fig_linewidth}, in which a balanced PD is used for suppressing the detectors' common noise. We use an AOM to frequency shift the whole comb by 55 MHz and heterodyne the comb with itself after sub-coherent delay. Since the delay length is less than the coherent length, the heterodyned optical signals are not entirely independent. The phase noise and linewidth are recovered by the following relation:
\begin{equation}
\label{linewidth}
    S_{\phi}(f)=S_{\Delta\phi}(f)\cdot\frac{1}{4\sin^{2}(\pi f\tau)}=\frac{\delta\nu}{\pi f^{2}}, 
\end{equation}
where $S_{\phi}$ is the phase noise of the comb laser, $S_{\Delta\phi}$ is the phase noise measured from the sub-coherent heterodyne measurement, and $\delta\nu$ is the Lorentzian linewidth. We performed the measurement with different delays from single-mode fibers with different lengths of 17 m, 30 m, and 1 km, and obtained the same intrinsic linewidth. The results shown in the main text are measured with a 30-m delay fiber.\\
\indent Noticeably, the measured linewidth represents the overall linewidth contributed from the entire comb, which indicates the upper limit of the intrinsic linewidth, as the entire comb is self-heterodyned without any filtering. We employ this method to more effectively showcase the overall coherence of the comb and to further substantiate the phase-locking among all comb lines.
\subsection{Fast frequency tuning of comb}
\begin{figure}
	\centering
	\includegraphics[width=\linewidth]{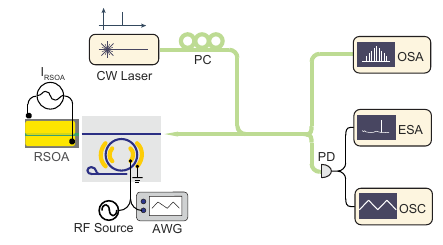}
	\caption{\label{tune_setup} Experimental setup for the comb tuning experiment. An arbitrary waveform generator (AWG) is used for modulating the comb laser. The comb output is beaten with an external CW laser at 1582~nm, and the real-time beating signal from PD is recorded by the OSC. Comb tuning is recovered from the post-processing of the beating signal. PC: polarization controller; OSC: real-time oscilloscope}
 
\end{figure}

In the experiment of high-speed comb frequency tuning described in the main text, the comb frequency modulation was measured by recording the temporal waveform of a beating signal between the comb and an external single-frequency CW laser near 1582~nm (see Fig.~\ref{tune_setup}). The external laser is tunable to control the beating node frequency, which was chosen to be 15~GHz as the carrier frequency in the experiment. When the triangular driving signal is turned off, the beating signal appears as a sinusoidal temporal waveform with a constant frequency of 15~GHz, similar to the trace in Fig.~2\textbf{h} of the main text. 
The triangular driving signal introduces chirping to the beating signal. We recorded the real-time beating signal and retrieved the time-frequency spectrum through the short-time Fourier transform (STFT). The results are shown in Fig.~4\textbf{b-d} of the main text.\\
\indent The time-frequency spectra exhibit certain distortions on its time-dependent waveforms, which primarily arise from two factors. First, the waveform of the driving electric signal is distorted by some extent from an ideal triangular waveform, due to the limited bandwidth of the RF amplifier used to boost its amplitude, as evident on the dashed curves in Fig.~4\textbf{b-d} of the main text. This part of waveform distortion is transferred directly to the comb frequency tuning waveform. Second, the time window used in the STFT introduces rounding and distortion particularly around the transition regions between frequency up-chirping and down-chirping. On the other hand, the time-frequency spectra also exhibit certain blurring on the waveform particuarly at a high modulation speed, which is simply due to the limited time window used for STFT. 

To determine the waveform deviation between the driving electrical signal and the comb frequency tuning, we find the spectral peak of a time-frequency spectrum at every time point, which forms the time-dependent frequency tuning curve of the comb. We compare it with the recorded time-dependent waveform of the driving electrical signal. For a fair comparison, rounding effect due to the limited time window is taken into account for the waveform of the driving electric signal. By comparing the two curves, we obtained the waveform deviation curve shown in the lower panels of Fig.~4\textbf{b}-\textbf{d} of the main text. 


\section{Theoretical modeling}
In the main text, we provide a comprehensive description of the design and experimental implementation. Here, we employ a simplified model to elucidate the underlying physical mechanisms of the comb laser operation.
Fig.~\ref{FigS1} provides the simplified schematics of the comb laser systems, including a conceptual overview of the key components for both the cavity-enhanced (CE) and Fabry-Perot (FP) comb lasers. In this section, we will delve into the system dynamics and aim to substantiate the experimental findings through mathematical analysis and numerical simulations.
\subsection{Dynamics of comb laser}
\begin{figure}[h]
	\centering
	\includegraphics[width=\linewidth]{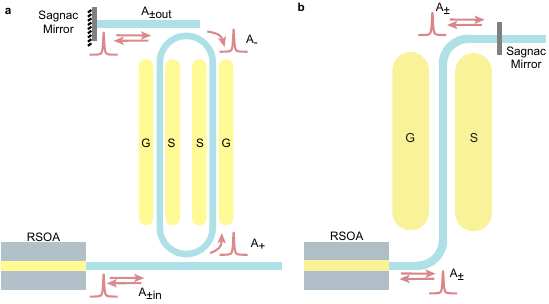}
	\caption{\label{FigS1} Device schematics. {\bf a,} Cavity-enhanced CE comb laser device consists of RSOA, EO modulated resonator, and broad-band Sagnac mirror. The optical field of the resonator and the lower/upper bus waveguide are represented by A, $\rm A_{in}$ and $\rm A_{out}$ respectively. {\bf b}, FP comb laser device, the laser cavity only includes RSOA, phase-shifting waveguide section, and Sagnac mirror. The optical field inside the laser is represented by A.}
\end{figure}

The comb lasing dynamics involves EO modulation and FWM process inside the laser cavity. Here we analyze the cavity-enhanced comb laser as an example. The FP comb laser can be analyzed using a similar approach by removing corresponding terms. The physical configuration of the cavity-enhanced laser is shown in Fig.~\ref{FigS1}\textbf{a}, highlighting its three primary components: the RSOA, the resonator, and the Sagnac mirror. The electrodes at the resonator adopt a Ground-Signal-Signal-Ground (GSSG) configuration and are modulating at one FSR of the racetrack, thereby introducing the coupling between adjacent resonant modes.
\begin{figure*}
	\centering
	\includegraphics[width=\linewidth]{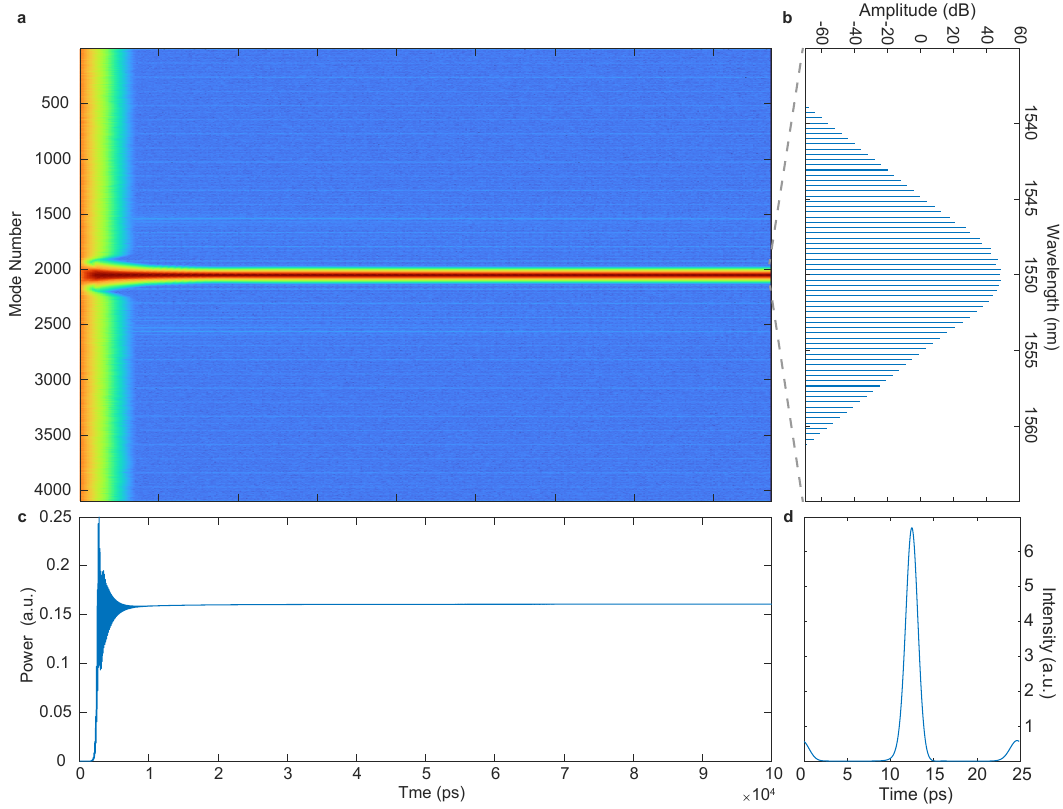}
	\caption{\label{FigS2_1} Simulation result for the turning-on process of a CE comb laser. The gain is turned on at time zero, with a constant EO driving frequency of one FSR. {\bf a,} Time-dependent spectrum evolution of the comb self-starting process. {\bf b,} Final spectrum of the lasing soliton comb. {\bf c,} Time-dependent intra-cavity power evolution of the comb self-starting process. {\bf d,} Final time domain pulse shape of the lasing soliton.}
\end{figure*}
\par To quantitatively analyze the dynamics, we use a model similar to the nonlinear Shr$\rm \ddot{o}$dinger equation (NLS) and the Lugiato–Lefever equation (LLE) and try to study the dynamics majorly in the time domain for the intracavity waveform of the racetrack resonator. We first consider the combined effect of Kerr and laser gain, and the EO modulation will be included next as an extra term for the modified LLE. The optical field at the resonator and the two bus waveguides can be written in $A_{\pm}$, $A_{\pm in}$ and $A_{\pm out}$ as shown in Fig.~\ref{FigS1}{\bf a}, where "+" corresponds to the counter-clockwise (CCW) wave and "-" represents the clockwise (CW) wave. Considering that the Sagnac mirror has high flat reflectance, the CW wave and CCW wave have no nonlinear interaction inside the cavity, we can assume that the CW wave and CCW wave are identical except for fixed-length beam propagation. Also, since the power inside the racetrack resonator is significantly higher than the main laser cavity, we simplify the model by neglecting the nonlinearity of the main laser cavity. Besides, we combine the nonlinear process of CCW and CW resonant modes and only study them in one direction. Symmetric coupling conditions ($\theta=\theta_{1}=\theta_{2}$) for the different bus waveguides to resonator coupling are also assumed in this analysis. As the optical field of the CW direction remains identical to that of the CCW direction plus beam propagation, we only study $A_+$ in the CCW direction and rewrite it as $A$, with the following relations:
\begin{align}
\nonumber
t_R\frac{\partial A(t,\tau)}{\partial t}=&\Big[-\alpha-i\delta_0+iL\sum_{k\geq 2}\frac{\beta_k}{k!}(i\frac{\partial}{\partial \tau})^k+ i\gamma L|A|^2\Big]A\\
\label{LLE_TD1}
&+\sqrt{\theta}A_{in},\\
\label{LLE_TD2}
A_{in}(t,\tau)=&G_R\cdot({1+iL_l\sum_{k\geq 2}\frac{\beta_{kl}}{k!}(i\frac{\partial}{\partial \tau})^k})\cdot A(t,\tau-t_{Rl}),\\
\label{LLE_TD3}
G_R=& e^{-\alpha_l}\cdot\theta^2\cdot R(\omega)\cdot \frac{g_m(\omega)}{1+|\bar{A}|^2/|A_{sat}|^2}.
\end{align} 
\begin{figure*}
	\centering
	\includegraphics[width=\linewidth]{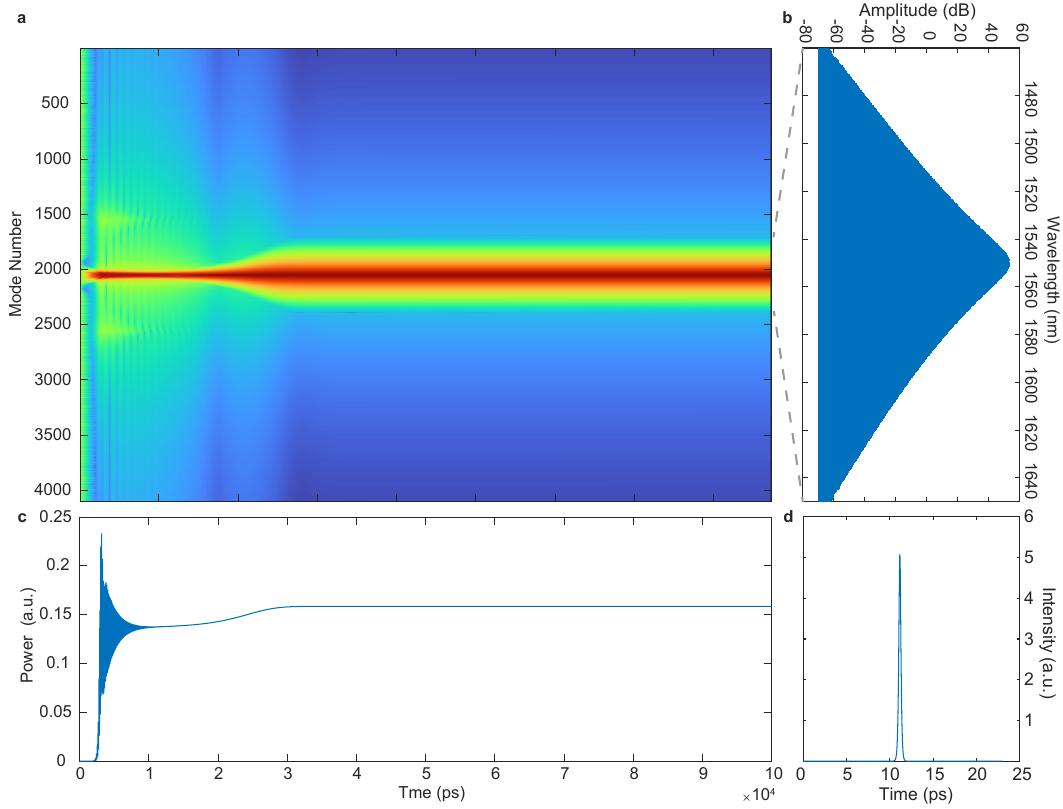}
	\caption{\label{FigS2_2} Simulation result for another example of turning on a CE comb laser, with stronger RSOA gain and significantly wider spectrum. {\bf a,} Time-dependent spectrum evolution of the comb self-starting process. {\bf b,} Final spectrum of the lasing soliton comb. {\bf c,} Time-dependent intra-cavity power evolution of the comb self-starting process. {\bf d,} Final time domain pulse shape of the lasing soliton.}
\end{figure*}
\if{ 
\begin{align}
	\label{LLE_TD_bydirection}
	t_R\frac{\partial A_+(t,\tau)}{\partial t}=&\Big[-\alpha+iL\sum_{k\geq 2}\frac{\beta_k}{k!}(i\frac{\partial}{\partial \tau})^k+ i\gamma L|A_+|^2\Big]A_++\sqrt{\theta}A_{+in}-\sqrt{\theta}A_{+out}(t,\tau+t_R/2)\\
	A_{+out}=&\sqrt{\theta}A_+(t,\tau+t_R/2)\\
	A_{-out}=&R\cdot({1+iL_l\sum_{k\geq 2}\frac{\beta_{ktop}}{k!}(i\frac{\partial}{\partial \tau})^k})\cdot A_{+out}(t,\tau-t_{top})\\
	t_R\frac{\partial A_-(t,\tau)}{\partial t}=&\Big[-\alpha+iL\sum_{k\geq 2}\frac{\beta_k}{k!}(i\frac{\partial}{\partial \tau})^k+ i\gamma L|A_-|^2\Big]A_-+\sqrt{\theta}A_{-out}-\sqrt{\theta}A_{-in}(t,\tau+t_R/2)\\
	A_{-in}=&\sqrt{\theta}A_-(t,\tau+t_R/2)\\
	A_{+in}=&G_R\cdot({1+iL_l\sum_{k\geq 2}\frac{\beta_{kbottom}}{k!}(i\frac{\partial}{\partial \tau})^k})\cdot A_{-in}(t,\tau-t_{bottom})\\
	G_R=& e^{-\alpha_l}\cdot\theta^2\cdot R(\omega)\cdot \frac{g_m(\omega)}{1+|\bar{A}|^2/|A_{sat}|^2}
\end{align}
}\fi
The physical process can be broken down into three equations. The Eq.~\ref{LLE_TD1} is the LLE equation of the racetrack resonator\cite{coen2013universal}.  Here, $A$ is the optical field inside the racetrack such that $|A|^2$ is normalized to optical power, and $A_{in}$ is the optical field inside the bus waveguide as the input for the resonator. The detuning $\delta_0$ is zeroed when the racetrack resonator central mode matches with the main laser cavity. The optical field change in the bus waveguide per main laser cavity roundtrip is discussed in Eq.~\ref{LLE_TD2} and Eq.~\ref{LLE_TD3}. For the bus waveguide of main laser cavity, the light coupled into and out of the racetrack twice per roundtrip, reflected once by the Sagnac mirror and amplified once by the RSOA, all of these beam propagation processes are written together as roundtrip gain $G_R$ in the Eq.~\ref{LLE_TD2}. Meanwhile, the dispersion of beam propagation of the bus waveguide is also included as the term
$iL_l\sum_{k\geq 2}\frac{\beta_{kl}}{k!}(i\frac{\partial}{\partial \tau})^k$. The roundtrip gain $G_R$, on the other hand, is discussed by Eq.~\ref{LLE_TD3}, and is related to loss, reflectance, and saturated RSOA gain.\\
\indent In these equations, $t_R$ and $t_{Rl}$ is the roundtrip time of the racetrack resonator and the laser main cavity, $L$ and $L_l$ is the normlized cavity length of the racetrack and the laser main cavity. $\alpha=\alpha_0/2+\theta_1/2+\theta_2/2$ is the total loss (where $\alpha_0$ is the intrinsic absorption coefficient and $\theta$ is the coupling coefficient between the resonator and the two bus waveguide), $\alpha_l$ is the loss of the laser main cavity.  $\beta_k$ and $\beta_{kl}$ is the averaged $k^{\rm th}$ order dispersion coefficient of the resonator and laser main cavity,  $\gamma = \frac{n_2\omega_0}{cA_{\rm eff}}$ is the nonlinear coefficient from Kerr effect. $G_R$ denotes the roundtrip gain of the laser, where R is the reflectance of the Sagnac mirror and RSOA, $g_m$ and $A_{sat}$ here are the small signal gain and the saturated optical field of RSOA. The wavelength dependency of gain $g_m(\omega)$ and reflection $R(\omega)$ are addressed as function, but the time response of RSOA is considered to be instantaneous in this simplified model.
\par The above equations cover the dynamics for passive mode locking and here we include EO modulation as a fast time-dependent phase modulation. When EO modulation occurs at the k-th order of the racetrack resonator's FSR, the modified LLE equation Eq.~\ref{LLE_TD1} for comb laser can be re-write as the following:
\begin{align}
\begin{split}
	\label{LLE_EO}
	t_R\frac{\partial A(t,\tau)}{\partial t}=&\Big[-\alpha-i\delta_0+iL\sum_{k\geq 2}\frac{\beta_k}{k!}(i\frac{\partial}{\partial \tau})^k+ i\gamma L|A|^2+\\ 
 & i\eta_{EO}E_0 cos({\frac{2k\pi\tau}{t_R}})\Big]A+\sqrt{\theta}A_{in},
 \end{split}
\end{align}
where $\eta_{EO}$ is the EO coefficient and $E_0$ is the electrical field strength, the combination of which forms the non-dimensional phase change. A simple model of cosine phase oscillation is used here to represent the EO modulation. After introducing the EO effect, the Eq.~\ref{LLE_EO}, Eq.~\ref{LLE_TD3} and Eq.~\ref{LLE_TD2} form a complete set for the theoretical description of the CE comb laser.
\subsection{Simulation of comb laser self-starting}
\par With the above theoretical model, the numerical simulation can be run using the split-step Fourier method. As can be seen in the Fig.~\ref{FigS2_1}. The figure shows the activation process of the laser with a constant EO driving signal of one FSR applied on the electrodes. The comb laser used in this simulation has an intrinsic optical Q of 1.5 million, coupling Q of 3 million, $\beta_{2}$ of the resonator and $\beta_{2l}$ of the main laser cavity are -0.02 $ps^2/m$ and 0.05 $ps^2/m$, respectively. The main laser cavity optical path is exactly 3 times the racetrack resonator's. \\
\indent As can be seen in Fig.~\ref{FigS2_1}{\bf a} and {\bf c}, when the laser gain is turned on, a chaotic comb is started at first but quickly settled to a stable state, which ends at the soliton states shown by Fig.~\ref{FigS2_1}{\bf b} and {\bf d}, with the noises greatly suppressed. From the power diagram, the laser power first climbs up and rings after turning on and quickly stabilizes to a constant value. The spectrum of the final stable single soliton state features a $sech^2$ shape with a bandwidth around 20 nm, as shown in Fig.~\ref{FigS2_1}{\bf b}. Such a shape is similar to the testing results of Laser $\beta$ in the previous section. The corresponding time domain pulse shape in Fig.~\ref{FigS2_1}{\bf d} is also consistent with the autocorrelation trace obtained experimentally. It's particularly noteworthy that, unlike the Continuous Wave (CW) pumped frequency comb that features a central residual comb line, the lased soliton comb in this system exhibits no CW background. This characteristic is evident in Fig.~\ref{FigS2_1}{\bf b} and {\bf d} and aligns well with the experimental results presented in the main text. The soliton comb extracts all of its energy directly from the RSOA, thereby maximizing conversion efficiency.\\
\indent Changing the RSOA gain or the maximum optical power output of RSOA, and fixing the EO driving, we can find another state of the CE comb laser output which features a significantly wider bandwidth. With increased optical power, the FWM expands the comb greatly, and the optical bandwidth of the comb increases to 150 nm in Fig.~\ref{FigS2_2}\textbf{b}, leading to a shorter time domain pulse in Fig.~\ref{FigS2_2}\textbf{d}. The time domain pulse shape also has a clean background and high extinction ratio, which is fundamentally different from normal Kerr cavity soliton.  Changing the optical power, we are also able to control the bandwidth of the output spectrum in the simulation just like the example in Fig.~5 of the main text, and all the initiation processes stay robust. Additionally, this suggests that enhanced coupling, increased RSOA strength, and refined roundtrip time control could extend the optical bandwidth of the comb laser far beyond the current limit.\\
\begin{figure}
	\centering
	\includegraphics[width=\linewidth]{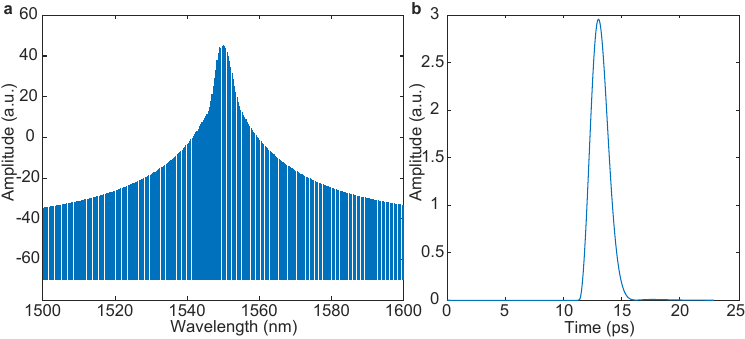}
	\caption{\label{FigS3} Simulation result for CE comb laser, with constant strong EO driving frequency at one FSR but low RSOA output power to suppress the FWM process. {\bf a,} The final spectrum of the lasing EO comb. {\bf b,} The final time domain pulse shape of the lasing EO comb.}
\end{figure}
\subsection{EO comb lasing}
\indent The lasing comb also differs from the pure EO comb\cite{zhang2019broadband} drastically. EO modulation creates correlations between different laser modes to start multimode lasing which in turn serves as a means to initiate and lock the comb lasing, instead of directly distributing power from the center line. To better demonstrate this, we decrease the RSOA output power to a negligible level (no FWM) but increase the EO modulation power to form a lased EO comb as shown in Fig.~\ref{FigS3}. Due to the low optical power, FWM barely participates in the process and the spectrum of the EO comb features a special shape shown in Fig.~\ref{FigS3}{\bf a}, with the corresponding pulse shape in Fig.~\ref{FigS3}{\bf b}. This EO comb laser resembles the traditional active mode-lock laser, where only EO modulation serves as the locking mechanism.\\
\bibliography{sup_reference}
\bibliographystyle{naturemag} %